\documentclass[aps,10pt,twocolumn,nofootinbib]{revtex4-2} 
\usepackage[T1]{fontenc}
\usepackage{mathptmx}
\usepackage{datetime}
\usepackage{amsmath}
\usepackage{amsfonts}
\usepackage{amssymb}
\usepackage{mathrsfs}
\usepackage[mathscr]{euscript}
\usepackage[dvips]{graphicx}
\usepackage{fancyhdr}
\usepackage{colordvi}
\usepackage{hyperref} 
\usepackage{epsfig}
\usepackage{color}
\usepackage{bm}
\usepackage{enumitem}

\usepackage{tikz}

\def\overstrike#1#2{{\setbox0\hbox{$#2$}\hbox to \wd0{\hss
    $#1$\hss}\kern-\wd0\box0}}

\newdateformat{yymmdddate}{\THEYEAR/\twodigit{\THEMONTH}/\twodigit{\THEDAY}}

\newcommand\qthinga[2]{{#1}_{#2}\vphantom{{#1}_{i}^{\dagger}}}
\newcommand\qthingd[2]{{#1}_{#2}^{\dagger}\vphantom{{#1}_{#2}^{\dagger}}}

\newcommand\aaai[1]{\qthinga{\alpha}{#1}}
\newcommand\aadi[1]{\qthingd{\alpha}{#1}}

\newcommand\oaaai[1]{\hat{a}_{#1}^{\vphantom{\dagger}}}
\newcommand\oaadi[1]{\hat{a}_{#1}^\dagger}

\def\kpump{u}
\def\ksigl{s}
\def\kidlr{i} 

\def\knlc{\vphantom{\chi^{*}}\chi}

\def\tinyzero{{\scalebox{0.6}{0}}}
\def\tinyhalf{{\scalebox{0.6}{1/2}}}

\def\hbarr{\,{\mathchar'26\mkern-9muh}}

\def\cRed#1{{{#1}}}

\begin{document}
\title{The surprising persistence of time-dependent quantum entanglement}
\author{Paul Kinsler}
\homepage[]{https://orcid.org/0000-0001-5744-8146}
\email{Dr.Paul.Kinsler@physics.org}
\author{Martin W. McCall}
\homepage[]{https://orcid.org/0000-0003-0643-7169}
%
\author{Rupert F. Oulton}
\homepage[]{https://orcid.org/0000-0002-5070-3623}
%
\author{Alex S. Clark}
\homepage[]{https://orcid.org/0000-0002-9161-3667}
%
%
\affiliation{
  Department of Physics,
  Imperial College London
  United Kingdom
}

\begin{abstract}








The \cRed{mismatch between elegant}
 theoretical models and 
 the detailed experimental reality 
 is particularly pronounced in
 \cRed{quantum nonlinear interferometry (QNI)}.
\cRed{In stark contrast to theory,
 experiments contain} 
 pump beams that start in impure states and that are depleted, 
 quantum noise that affects -- and drives -- 
 any otherwise gradual
 build up of the signal and idler fields, 
 and nonlinear materials that are far from ideal and
 have a complicated time-dependent dispersive response.
\cRed{Notably, 
 we would normally expect group velocity mismatches to 
  destroy any possibility of measurable or visible entanglement,
  even though it remains intact -- 
  the mismatches change the relative timings 
  of induced signal-idler entanglements, thus generating ``which path''
  information.}
 \cRed{Using a ``positive-P'' approach
 ideally suited to such problems,
 we show how 
 the time-domain entanglement 
 crucial for
 QNI can be --
 and is --
 recoverable despite
 the obscuring effects of 
 real-world complications.}

\end{abstract}


\date{\today}
\maketitle

%

%

Quantum entanglement is important because it
 plays a key role in a range of quantum devices, 
 notably in induced coherence \cite{Zou-WM-1991prl,Wang-ZM-1991pra,Wiseman-M-2000pla}
 based quantum imaging / QNI schemes 
 \cite{Viswanathan-LL-2021arxiv,Fuenzalida-HLLLZ-2020arxiv,Lahiri-HLLZ-2019pra,
Moreau-TGP-2019nrp,Padgett-B-2019prsta,Basset-SSBPG-2019lpr,Chekhova-O-2016aop,KOBQI}; 
 and 
 in the time domain is a subject of wide-ranging and active study 
 \cite{Yang-ZDPLZCJ-2021pra,Aumann-PKODPRLW-2021arxiv,De-GBSSHR-2021arxiv,Tran-N-2021arxiv,Castellani-2021arxiv}.
However, 
 the complicating effects of material dispersion
 in the entanglement-generating nonlinear medium, 
 or during subsequent propagation,
 are typically not considered \cite{Marcikic-RTZLG-2004prl,Inagaki-MTAT-2013oe,Yu-MLJSFYetc-2020n}.
This has wider relevance,
 not only for quantum interference in general,
 but also e.g.
 in quantum data transmission \cite{Pirandola-LOB2017nc,Zwerger-PDBD-2018prl,Khatri-CSD-2019prl}
 and QNI.
We demonstrate in this Letter how and why, 
 \cRed{despite the complete 
 de-synchronization of entangled fields
 caused by material dispersion, 
 a slow detection process
 can perform an unexpected ``entanglement recovery'', 
 so that time resolved QNI
 can, 
 after all,
 unexpectedly succeed.}

Our testbed for examining the 
 limitations on time-resolved quantum measurements
 is a pulsed QNI
 system where 
 time dependence is relevant for all field interactions,
 in particular with the material dispersion 
 (i.e. group velocity, and group velocity dispersion (GVD))
 present alongside the entanglement-generating nonlinearity.
In ghost imaging, 
 for example, 
 one can imagine a clear separation between standard (spatial)
 schemes \cite{Pittman-SSS-1995pra,Erkmen-S-2010aop}
 and temporal schemes \cite{Shirai-SF-2010josab,Ryczkowski-BFDG-2016nphot}, 
 but if material dispersion was present during propagation, 
 such simplicity would be disrupted.
To address such intrinsic complications
 requires a shift in both theoretical methods and mindset; 
 a description can no longer rely on using only a small number 
 of possible states (typically Fock states),
 and judgements based on path indistinguishability
 or phase shifts.
Instead, 
 a 
 set of time dependent states
 is required, 
 {and consequently it is not path lengths 
 but relative timings that matter.}

\begin{figure}
\includegraphics[width=0.75\columnwidth]{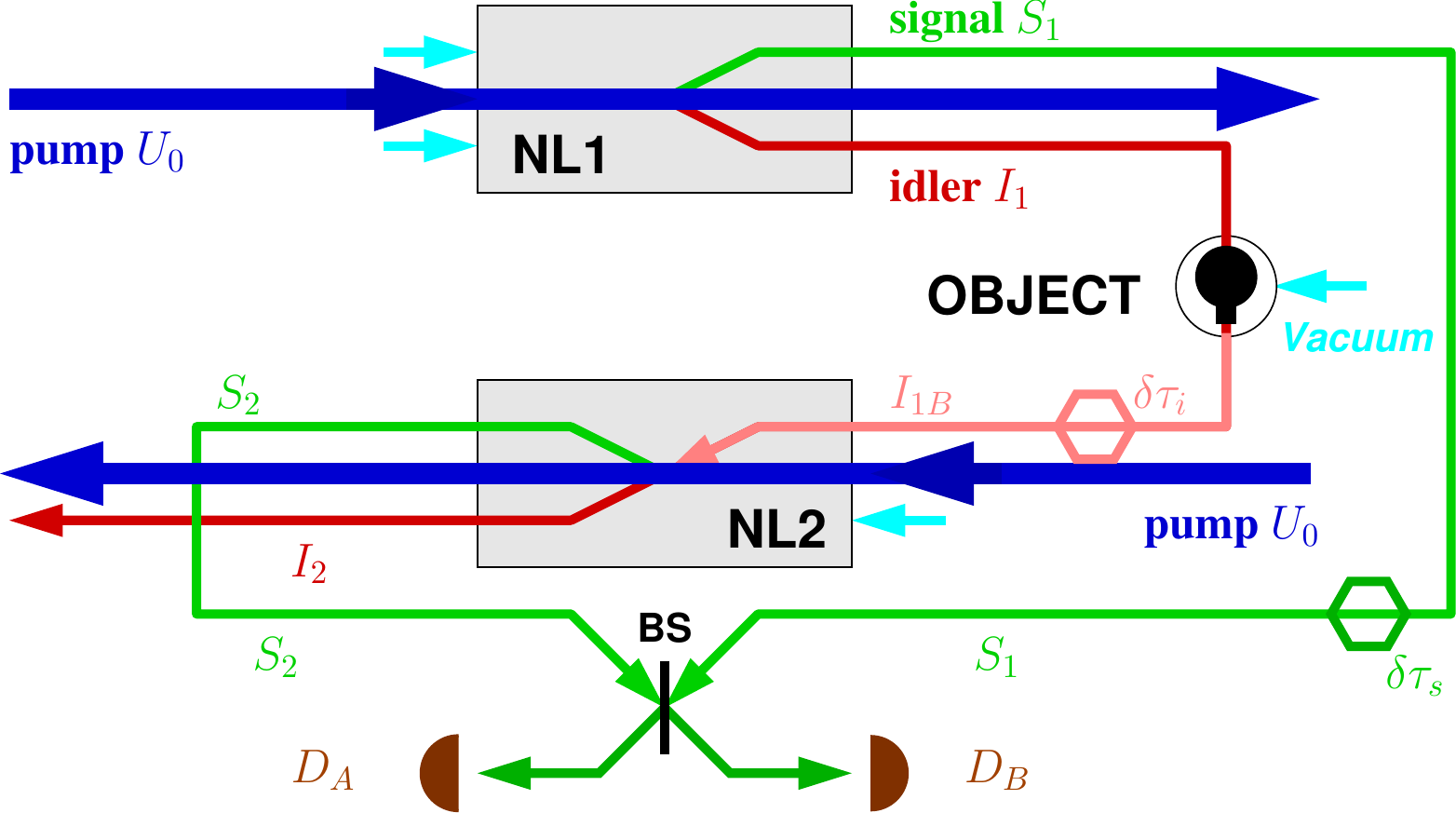}
\caption{A representation of the 
 QNI scheme (e.g. \cite{Kolobov-GLFB-2017jo}).
The two identical pump fields are in blue, 
 signal fields in green, 
 and idler in red; 
 here the FWM process combines
 two pump photons to create a signal-idler pair.
The resulting signal fields then \cRed{interfere} at the beamsplitter (BS)
 and detected at $D_A$ and $D_B$.
In an Imperial experiment \cite{Pearce-POC-2020apl},
 the nonlinear stages (NL1 and NL2)
 take place in opposite directions
 through the same length of fibre.
}
\label{fig-schematic}
\end{figure}

How the 
 fields propagate and are transported
 through the QNI system
 is shown on figure \ref{fig-schematic}; 
 the layout is very similar to that of
 Kolobov et al \cite{Kolobov-GLFB-2017jo}. 
The key feature of the nonlinearity is that 
 it produces correlated signal and idler photon pairs 
 from an incoming pump field; 
 this is often achieved using a $\chi^{(2)}$ interaction, 
 but here we use a  {degenerate-pump} four wave mixing (FWM) 
 (see e.g. \cite{Pearce-POC-2020apl}).
A pump field enters the first nonlinear stage (NL1)
 and generates entangled signal and idler fields; 
 and whilst the signal is diverted to the final beamsplitter,
 the idler instead interacts with the to-be-imaged object,
 and
 then serves as a co-input, 
 with a {copy of the original} pump field, 
 to the second nonlinear stage (NL2).
The signal field departing NL2
 is thus influenced by 
 an idler field entangled with the
 first signal field, 
 and this information is extracted
 by interference at the beamsplitter, 
 before {photon detection}.

\def\Pump#1{U_{#1}}
\def\Signal#1{S_{#1}}
\def\Idler#1{I_{#1}}
\def\Sigsav#1{R_{#1}}

The nonlinear propagation model in our simulations
 is a well-established one derived originally for the prediction
 of squeezing generation in optical fibres
 \cite{Drummond-C-1987josab,Carter-DRS-1987prl,Carter-1993phd,Carter-1995pra,Corney-HDJDLA-2008pra},
 has also been used to model multi-field parametric processes
 \cite{Werner-D-1993josab}, 
 and here centres primarily on multi-field nonlinear propagation
 through a dispersive material.
It also includes stages representing setting up the initial conditions, 
 interaction with an object to be imaged, 
 and mixing at a beamsplitter at the interferometer output.
In this work we use an established off-diagonal coherent state basis 
 positive-P approach
 \cite{Drummond-G-1980jpa,GardinerHSM,DrummondHilleryQTNO}, 
 that enables both group velocity and dispersion
 to be easily implemented in a numerical scheme \cite{Drummond-C-1987josab}; 
 crucially, 
 its off-diagonal nature allows a complete representation 
 of the full quantum mechanical density matrix of the system, 
\cRed{and of its dynamics.}

%

\def\kzero{k_{\tinyzero}}
\def\kmzero{k_{m\tinyzero}}
\def\omegazero{\omega_{\tinyzero}}
\def\omegamzero{\omega_{m\tinyzero}}

Our description
 uses time-dependent
 \cRed{quasi-conjugate} pump field amplitudes $\aaai{\kpump}(t)$, $\aadi{\kpump}(t)$, 
 \cRed{which are only complex-conjugate on average;}
 likewise for signal $\aaai{\ksigl}(t)$, $\aadi{\ksigl}(t)$, 
 and
 idler $\aaai{\kidlr}(t)$, $\aadi{\kidlr}(t)$.
There is 
 also a set of independent time-dependent noise increments $\{dW_a(t)\}$.
In simulation, 
 time is discretized \cRed{and labels a set of finely divided, 
 sub-pulse length modes where}
\cRed{each field is 
 held as an array of sequential ``time-bins'' at $t \in \{t_j\}$, 
 each of which contains pairs of complex field amplitudes; 
 these interact and evolve
 as the fields propagate step-by-step through space.
Since these time-bins have a simulation-specific duration,
 field averages such as $<\!\!\aadi{\ksigl}\aaai{\ksigl}\!\!>$
 represent intensity (in photons per second), 
 not photon number, 
 and normalizations and 
 parameter 
 scalings reflect this.}

This field information is held 
 along with fixed parameters for the 
 nonlinear coupling $\kappa$, 
 losses $\gamma_m$, 
 and dispersive properties $D_m(\omega)$, 
 where the field subscripts, 
 as above, 
 are $m \in \{\kpump, \ksigl, \kidlr, r\}$.
Since this is a stochastic technique, 
 very many independent copies of the evolution
 need to be run 
 (for the simulations here, 
 typically in the range $10^5$ to $10^7$),
 and the necessary ensemble averages taken.

%

This propagation model is very close to
 previous ones (e.g. 
  \cite{Drummond-C-1987josab,Carter-1995pra,Werner-D-1993josab}), 
 but here we have three co-propagating fields 
 and a different nonlinear interaction; 
 namely one for
 the degenerate FWM process with
 only resonant wave mixing terms
 where $2\omega_{\kpump} = \omega_{\ksigl} + \omega_{\kidlr}$.
Since self-phase modulation (SPM) terms are not significant
 in the low-power regime (see e.g. \cite{Pearce-POC-2020apl}, 
 the interaction Hamiltonian we need {here} is simply
~
\begin{align}
  \hat{H}_{\textup{fwm}}
&=
  \imath
  \hbarr
  {\knlc} 
  \oaadi{\kpump}\vphantom{a}^2 \oaaai{\ksigl}  \oaaai{\kidlr}  
 -
  \imath
  \hbarr
  {\knlc}^*
  \oaaai{\kpump}\vphantom{a}^2 \oaadi{\ksigl}  \oaadi{\kidlr}  
.
\label{eqn-dfwm-hamiltonian}
\end{align}
This interaction term results in propagation equations
 which are best expressed in the incremental form 
 suited to such stochastic differential equations (SDEs)
 \cite{Drummond-G-1980jpa}.
In an appropriate co-moving frame, 
 including loss along with nonlinear effects
 \cite{Carter-1995pra}, 
 but suppressing the time argument   
 common to all fields $\aaai{m}, \aadi{m}$
 and \cRed{quantum} noise increments $dW_a$,
 we have
~
\begin{align}
  d\aaai{\ksigl}
&= 
  \left[
   -
    \gamma_{\ksigl}
    \aaai{\ksigl}
   +
    {\knlc}^*
    \aaai{\kpump}
    \aaai{\kpump}
    \aadi{\kidlr}
  \right]
  dz
 ~~
 +
  \left\{
    (2{\knlc})^{\tinyhalf}
    \aaai{\kpump}
  \right\}
  dW_1
,
\label{eqn-cht-daa1}
\\
  d\aadi{\ksigl}
&= 
  \left[
   -
    \gamma_{\ksigl}
    \aadi{\ksigl}
   +
    {\knlc}
    \aadi{\kpump}
    \aadi{\kpump}
    \aaai{\kidlr}
  \right]
  dz
 ~~
 +
  \left\{
    (2{\knlc}^*)^{\tinyhalf}
    \aadi{\kpump}
  \right\}
  ~ dW_2
,
\label{eqn-cht-dad1}
\\
  d\aaai{\kidlr}
&= 
  \left[
   -
    \gamma_{\kidlr}
    \aaai{\kidlr}
   +
    {\knlc}^*
    \aaai{\kpump}
    \aaai{\kpump}
    \aadi{\ksigl}
  \right]
  dz
 ~~
 +
  \left\{
    (2{\knlc})^{\tinyhalf} \aaai{\kpump}
  \right\}
  ~ dW_1
,
\label{eqn-cht-daa2}
\\
  d\aadi{\kidlr}
&= 
  \left[
   -
    \gamma_{\kidlr}
    \aadi{\kidlr}
   +
    {\knlc}
    \aadi{\kpump}
    \aadi{\kpump}
    \aaai{\ksigl}
  \right]
  dz
 ~~
 +
  \left\{
    (2{\knlc}^*)^{\tinyhalf} \aadi{\kpump}
  \right\}
  ~ dW_2
,
\label{eqn-cht-dad2}
\\
  d\aaai{\kpump}
&= 
  \left[
   -
    \gamma_{\kpump}
    \aaai{\kpump}
   -
    2
    {\knlc}
    \aadi{\kpump}
    \aaai{\ksigl}
    \aaai{\kidlr}
  \right]
  dz
 ~~
 +
  \imath
  \left\{
    2 {\knlc}^{\vphantom{*}} \aaai{\ksigl} \aaai{\kidlr}
  \right\}^{\!\!\tinyhalf}
  \!\!\! dW_3
,
\label{eqn-cht-daap}
\\
  d\aadi{\kpump}
&= 
  \left[
   -
    \gamma_{\kpump}
    \aadi{\kpump}
   -
    2
    {\knlc}^* \!
    \aaai{\kpump}
    \aadi{\ksigl}
    \aadi{\kidlr}
  \right]
  dz
 +
  \imath
  \left\{
    2 {\knlc}^*  \!\!
    \aadi{\ksigl} \aadi{\kidlr}
  \right\}^{\!\!\tinyhalf}
  \!\!\! dW_4
.\!\!
\label{eqn-cht-dadp}
\end{align}
These
 equations are used to update each time-bin
 in the temporal profile of the fields
 as they propagate (step) forward in space.
Deterministic evolution terms are in square brackets $[...]$, 
 and prefactors for stochastic (noisy) terms in braces $\{...\}$.
The noises are uncorrelated, 
 with 
 $\delta_{ab} = < dW_{a}(t) dW_{b}(t) >$.

Here we see that there are both coherent 
 interactions between the three fields, 
 and correlated nonlinear quantum noise terms.
\cRed{The noise increment $dW_1$ 
 drives both $\aaai{\ksigl}$ and $\aaai{\kidlr}$,
 whilst $dW_2$ drives $\aadi{\ksigl}$ and $\aadi{\kidlr}$, 
 ensuring the pairs are correlated
 but \emph{not} complex conjugate.
The
 comparable classical model 
 (or even a semi-classical model,
  see e.g. \cite{Kinsler-D-1991pra,Kinsler-D-1991pra-comment,Kinsler-1996})
  would have
 only three equations
 and \emph{no} noise terms.}

Material dispersion is the other key feature, 
 and we interleave it with the nonlinearity in a split-step scheme
 (see e.g. Carter et al.\cite{Drummond-C-1987josab,Carter-1995pra}),
 using linear phase shifts 
 in the spectral domain
 for group velocity,
 and quadratic shifts for GVD.

%

\emph{Material parameters in the simulations}
 are chosen to be 
 compatible with fiber-based photon pair
 sources based on spontaneous four-wave mixing
 \cite{Pearce-POC-2020apl}, 
 which use about $100$cm of Thorlabs PM780-HP fibre
 (see Table \ref{table-params}).
For the simulation, 
 we convert parameters into units based on
 meters (m), 
 picoseconds (ps), 
 and field excitation amplitudes 
 referenced back to photon numbers per picosecond.
A crucial step here is to 
 consider pulsed operation 
 on a picosecond scale, 
 where the 
 group velocity mismatches are significant.
This regime is where the time-dependent nature
 of the entanglement will be most exposed to the 
 disruptive effects of material dispersion.

Since
 the effect of group velocity mismatches and GVD turns 
 localised entanglements into temporally distributed ones 
 in our simulations we see 
 a fan of signal amplitudes that 
 spreads out behind the pump pulse, 
 whilst a fan of idler amplitudes 
 spreads out before.
Thus, 
 crucially, 
 the signal-idler entanglement 
 is not only distributed over a range of times, 
 it is also between \emph{different} times; 
 i.e. it is a multi-time correlation.
For best imaging visibility, 
 this requires careful synchronisation at NL2, 
 where
 the first-generated part of the idler in NL2
 should be coincident with the pump pulse as it enters. 


\begin{table}[h]
 \begin{tabular}{|l|  c |  c |  r |  c |}  
 \hline
 ~Field~~    & ~Wavelength~ & ~~Frequency~~ & $\Delta v_g$ ~~~   &   ~Dispersion $d_2$~ \\ 
 \hline
 ~Pump     & 768 nm &   390.5 THz        &  0 ps/m            & 0.589 ps$^2$/km \\
 ~Signal   & 700 nm &   428.0 THz        &  -100 ps/m         & 0.489 ps$^2$/km \\
 ~Idler    & 850 nm &   353.0 THz        & +83 ps/m           & 0.721 ps$^2$/km \\
 \hline
\end{tabular}
\caption{Material parameters used in the simulations, 
 based on the experimental setup of
 \cite{Pearce-POC-2020apl}.
Since the fibre is weakly guiding, 
 these are based on those for bulk silica.
The nonlinearity in SI units 
 is $n_2 = 3 \times 10^{-20} $m$^2/$W,
 and the loss is $\gamma = 0.004/$m.
In simulation units, 
 the PM780-HP's stated transverse field mode area of about 25$\mu$m$^2$
 and
 the pump photon energy of $25 \times 10^{-20}$J, 
 means that
 the rescaled nonlinearity is $n''_2 = 0.3 \times 10^{-15}$
 per photon-picosecond. 
In a time-bin $T$ ps long,
 the pump photon has a power $\sim (0.25 / T)$ $\mu$W,
 so that
 a power of $100$W
 corresponds to  a flux of $4\times 10^8$ photons/ps. 
 Estimates for signal and idler fields are similar.
Note that 
 the pump powers used in the simulations are increased
 to enable good simulation statistics
 (see Supplementary information).
}
\label{table-params}
\end{table}

%

\emph{Objects} placed in the idler beam leaving the NL1 stage
 disrupt the entanglement with the signal beam, 
 and that disruption changes the numbers of detected photons, 
 enabling the object's presence or properties to be inferred.
However, 
 in non-imaging contexts,
 we can view them as representatives of 
 further disruptive unwanted real-world effects:
 loss, 
 phase shifts through optical elements, 
 or extraneous couplings.
\cRed{Here we consider 
 passive objects that impart 
 (a) a phase modulation $\Delta\phi$ of the idler field, 
 or have
 (b) a reflectivity $r$
 that reduces the idler amplitudes as they pass
 (as in  \cite{Kolobov-GLFB-2017jo}); 
 so $r=1$ removes all entanglement.}
We also consider (c) imaging of 
 dynamic objects, 
 a key feature 
 since an object's time dependence will
 affect visibility, 
 just as timing, group velocity, and GVD do.

Our linearly coupled dynamic objects
 \cRed{with amplitudes $\beta(t), \beta^\dagger(t)$}
 respond to the incident
 pulse profile $\alpha_m(t), \alpha_m^\dagger(t)$ 
 at the position $z$ specific to the object; 
 and a field-object interaction strength $\eta$.
At the object 
 we set the initial conditions at $t_o$
 so that $\beta(t_o)=\beta_o$ and $\beta^\dagger(t_o)=\beta^*_o$, 
 and then \emph{time} integrate $\beta(t), \beta^\dagger(t)$ using
~
\begin{align}
  d\beta(t)  
&=
  \left[ - \gamma_o \beta(t) + \eta^* \alpha_m(t) \right] 
  ~dt,
\label{eqn-dynobject-b}
\\
  d\beta^\dagger(t)  
&=
  \left[ -\gamma_o \beta^\dagger(t) + \eta^{~} \alpha_m^\dagger(t) \right] 
  ~dt
.
\label{eqn-dynobject-bdagger}
\end{align}
Now that the incident field $\alpha_m, \alpha_m^\dagger$
 has excited the dynamic object, 
 this excitation acts back on the field
 and modifies it.
We therefore then update $\alpha(t), \alpha^\dagger(t)$ 
 according to the same linear interaction 
 but here integrated forward in \emph{space}, 
 using
~
\begin{align}
  d\alpha_m(t) 
&=
 -
  \tfrac{1}{2}
  \eta^{~} \beta(t) 
  ~dz;
\quad
  d\alpha_m^\dagger(t) 
=
 -
  \tfrac{1}{2}
  \eta^* \beta^\dagger(t) 
  ~dz
,
\label{eqn-dynobject-backaction}
\end{align}
 and keeping the SDE notation for consistency.
Note that this could be extended to allow for 
 nonlinear couplings or dynamics with the object,
 or to even use (e.g.) a two-level-atom or Raman models
   (see e.g. a classical counterpart \cite{Kinsler-N-2005pra}).

%


\emph{Detection and Visibility:}
The photon number rate 
 measured at the 
 detectors is taken to be the ensemble-averaged photon number
 of the relevant field at that point.
An improved detector model 
 could be implemented 
 using similar approach to time-dependent objects.
An important quantity is the \emph{visibility} of the entanglement,
 which is the 
 {difference} of the detector counts (``signal'')
 divided by the
 {sum} of the detector counts (``background'');
 \cRed{i.e. $|n_A-n_B| / (n_A+n_B)$.}
To suppress sampling artifacts 
 \cRed{in the pulse wings}
 we add to the background
 an offset
 of $0.05$\% of its maximum.

\begin{figure}
\includegraphics[angle=90,width=0.4500\columnwidth]{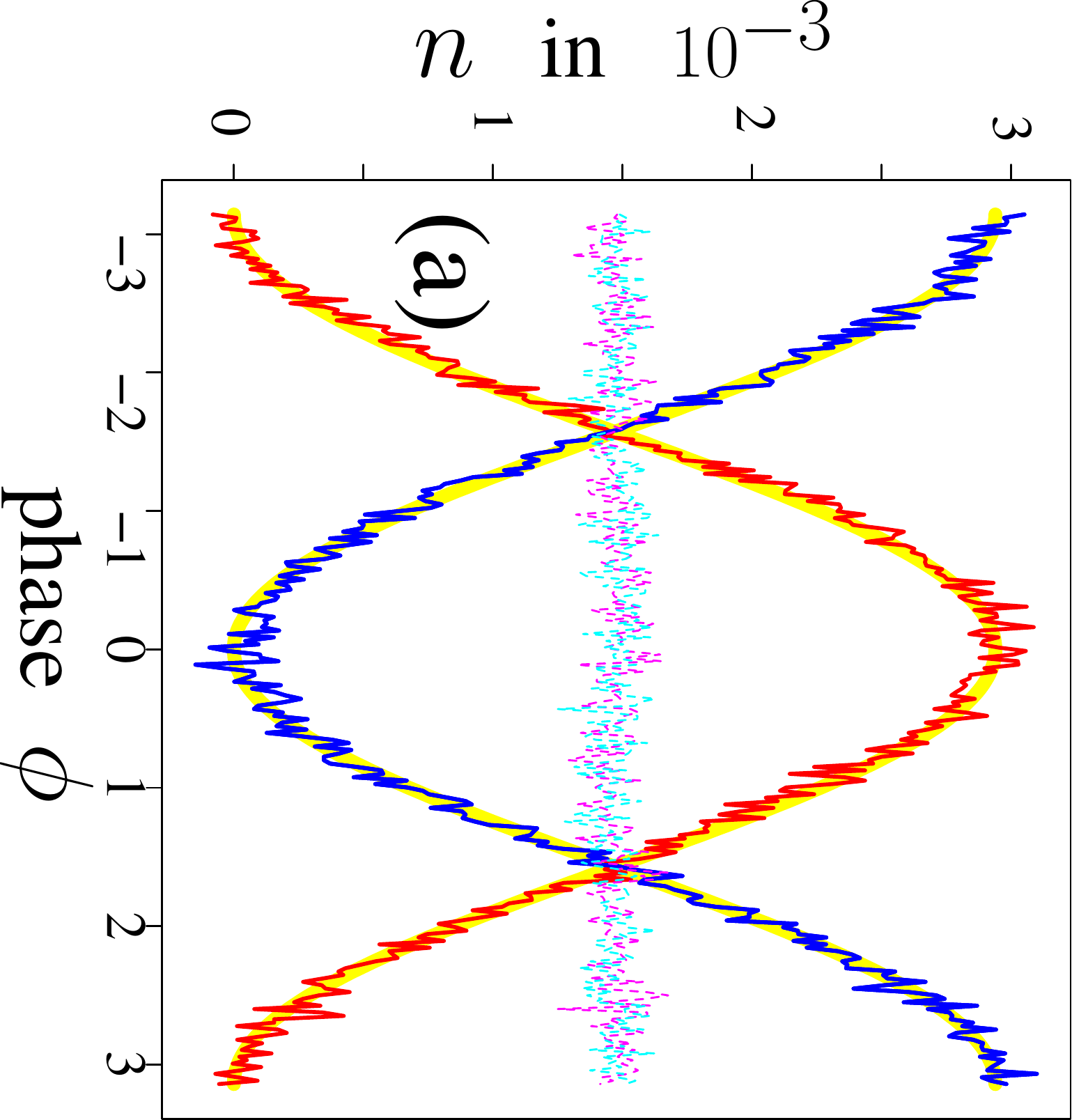}
\includegraphics[angle=90,width=0.4500\columnwidth]{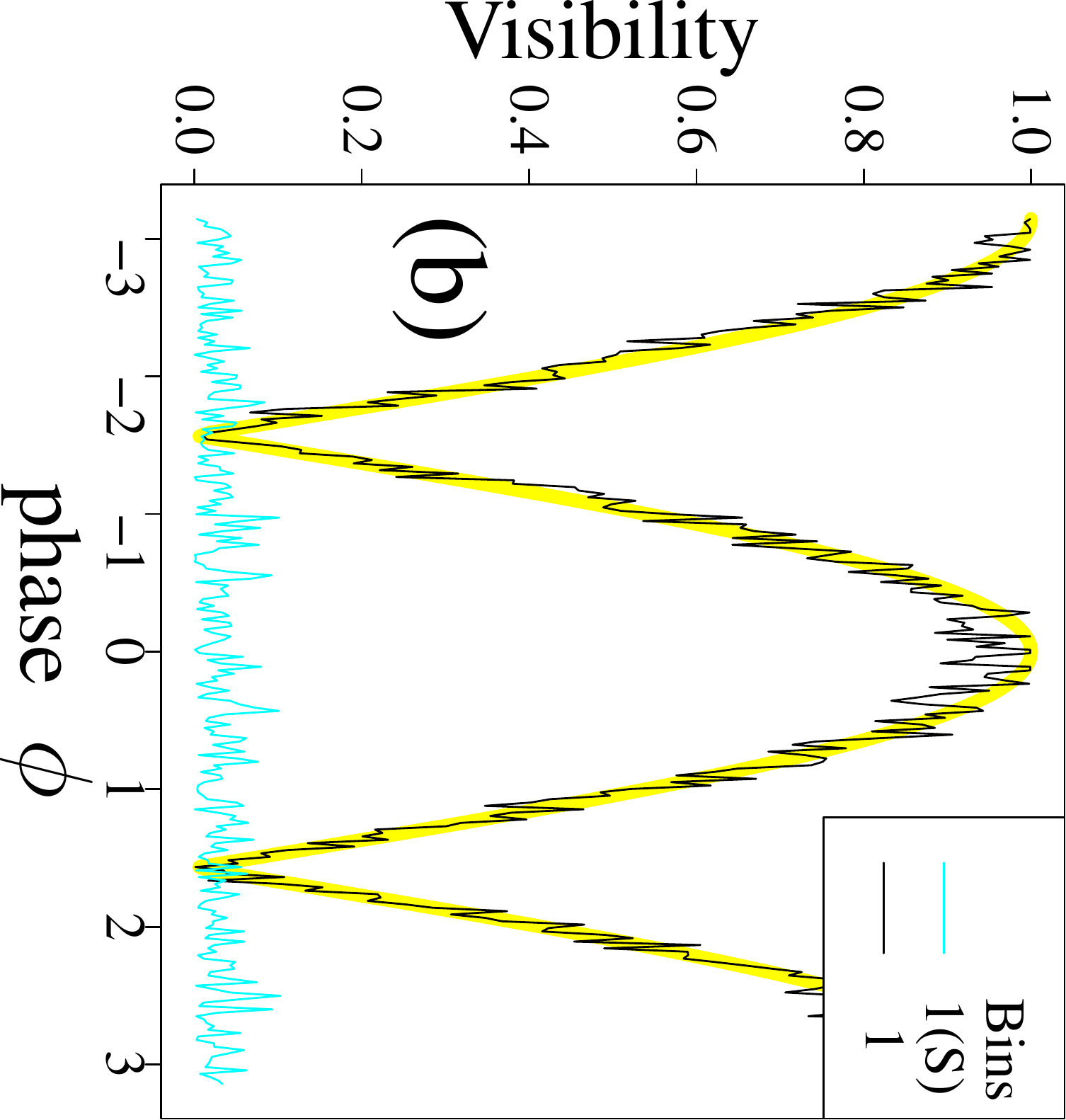}\\
\includegraphics[angle=90,width=0.4500\columnwidth]{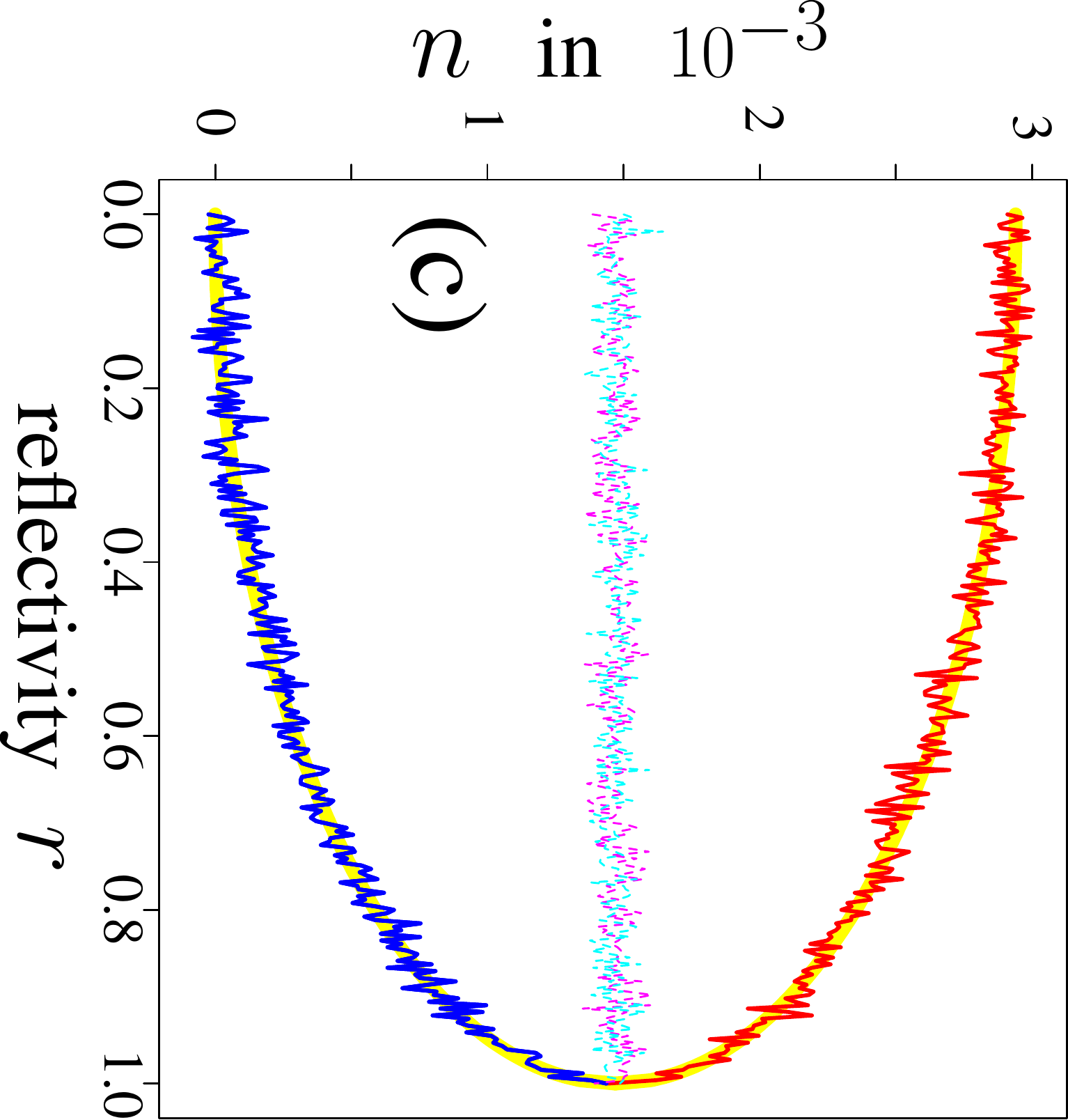}
\includegraphics[angle=-0,width=0.4500\columnwidth]{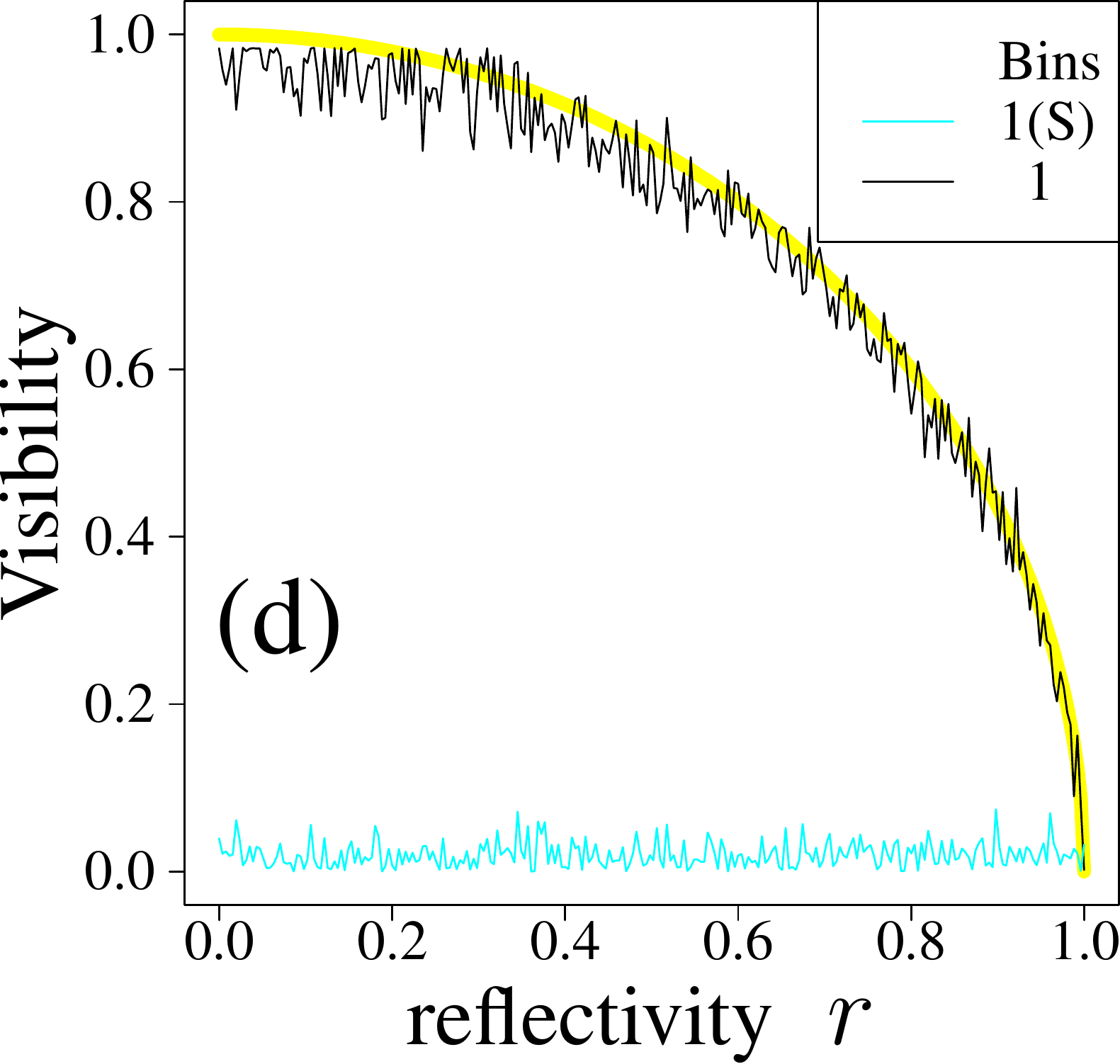}
\caption{\cRed{Entanglement} visibility measurement
 versus a specified phase shift (top) and reflectivity  (bottom)
 caused by an imaged object:
 \textbf{(a,c)} Photon counts at detectors $D_A$, $D_B$ (in red and blue),
   compared to photon numbers in the pre-beamsplitter signal fields
   $S_1$, $S_2$
   (dashed, in cyan and magenta).
 \textbf{(b,d)} 
   visibility based on detector measurements (black) 
   compared against one calculated from signal fields (cyan).
\cRed{On (c,d), the $r=1$ end result gives the case of a blocking object, 
 i.e. no entanglement.} 
\cRed{Statistical noise}
 can be reduced by increasing the ensemble sizes
 from 256k, 
 but are retained as-is to emphasise the 
 necessity of the ensemble averaging.
The results match 
 the theoretical fits shown in the background
 as thick yellow lines.
}
\label{fig-ideal-with-both} 
\end{figure}

We test the basics of the simulations
 with a simple CW-equivalent parameter set
 with no group velocity or GVD effects, 
 and look at how the visibility varies with 
 object phase-shifts and object reflectivities.
Here,
 correlations between the 
 first signal's time-bins and the second signal's ones 
 will always be synchronised, 
 maximizing the visibility.
For sufficiently low signal-idler generation efficiencies, 
 we should see equal photon number intensities in the signal's field-modes, 
 but distinct photon number intensities after the beamsplitter, 
 i.e. at the two detectors; 
 this is clearly shown on fig. \ref{fig-ideal-with-both}.

%

%

\begin{figure}
\includegraphics[angle=-0,width=0.450\columnwidth]{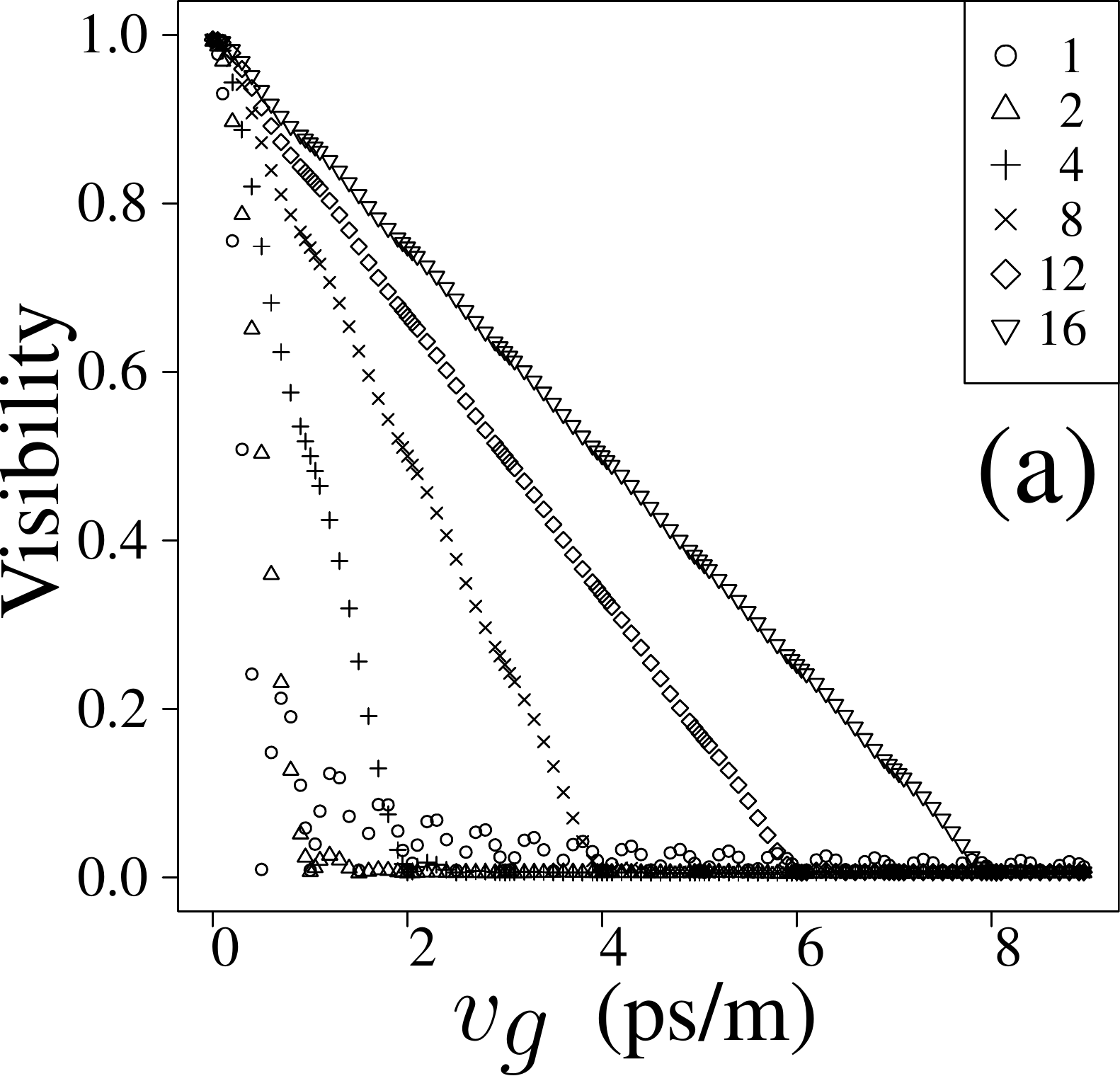}~~
\includegraphics[angle=-0,width=0.450\columnwidth]{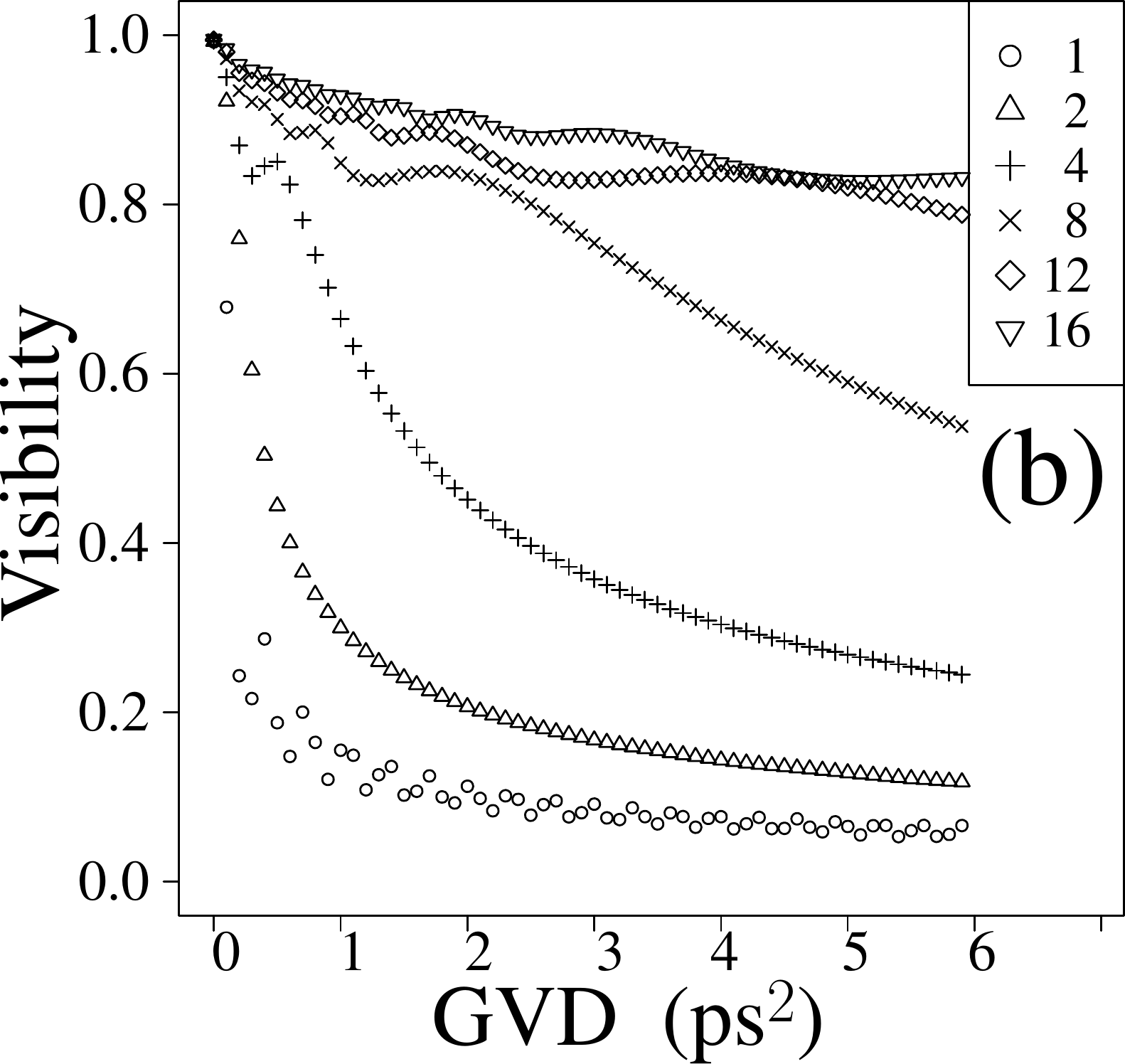}
\caption{\cRed{\emph{Detectable} entanglement is obscured
 by material dispersion, even though the entanglement itself remains:
 the no-object case, 
 as group velocity ($v_g$) and GVD mismatches increase,
 for 100cm propagation with standard parameters,
 and comparing detection time-averaging intervals.}
Here the mismatch is applied to the signal field $S_1$, 
 so that a $v_g$ mismatch 
 causes a timing offset $\delta \tau_s$ at the beamsplitter.
\cRed{Circles ($\circ$) are ``fast detector'' un-averaged results,} 
 other symbols show the number of adjacent $0.5$ps time-bins averaged.
For $v_g$ mismatches \textbf{(a)}, 
 the regular residual bumps in $\circ$
 are a discretization artifact, 
 and can be 
 further suppressed by increasing the time resolution.
The complicated 
 effect of GVD mismatches \textbf{(b)}
 leads to smooth variations
 as different contributions de-phase or rephase.
}
\label{fig-ideal-with-vgaveraging}
\end{figure}

Time averaging is a crucial part of 
 any detection model used here, 
 since real detectors are very slow
 (typically $\gg 100$ps)
 when compared to the temporal resolution of our simulations ($\sim$ps).
Although more sophisticated models can easily be imagined, 
 here we simply sum the time-binned amplitudes of each field
 over some chosen $m$-bin detector response time,
 before ensemble averaging to get the detected photons:
~
\begin{align}
 \bar{n}_{D} &= < \bar{\alpha}^{\dagger} \bar{\alpha}^{\vphantom{\dagger}} >,
 \quad  \textrm{with} ~~
  \bar{\alpha}^{(\dagger)}
=
  {(\tfrac{1}{m})}^{\!\tinyhalf}
  {\sum}_{i=1}^{m} \alpha^{(\dagger)}(t_i)
.
\label{eqn-detectorsum}
\end{align}
Crucially, 
 this averaging process in the detector helps expose
 correlations between time-bins that have become offset 
 due to dispersion mismatches.
{Note that this summation 
 is essentially the same as the process for
 combining multiple short time bins into a longer one; 
 just as we need to do in numerical convergence checks.}
On fig. \ref{fig-ideal-with-vgaveraging}, 
 \cRed{the entanglement is always fully present,
 but has its visibility reduced
 by $v_g$ or GVF mismatch.
However
 visibility can be recovered
  using longer averaging intervals; 
 at least up to}
 a ($v_g$) cut-off when the time difference 
 exceeds the averaging windows.


%

\emph{Pulsed simulations:}
Here we standardise on an input 40ps pump pulse
 and a 512ps window
 divided into $2$ps bins.
Simulation pulse intensities were chosen as low as practicable, 
 given the constraints of computation time 
 and the requirement for good simulation statistics.
Further, 
 the pump-idler pulses were ideally synchronised 
  as they entered the NL2 stage. 
However, 
 the signal fields are also mis-timed at the output beamsplitter
 by $\delta \tau_s = {20}$ps (i.e. {10} time bins).
This not only 
 mimics imperfect experimental setup, 
 it is also useful in providing an example where the detector averaging
 has more to recover.
We also use a technique,
 described in the Supplementary Material,
 to reduce the effect of sampling error --
 something which would otherwise make positive-P simulations either problematic
 or computationally prohibitive.

%

\begin{figure}
\includegraphics[angle=0,width=0.32450\columnwidth]{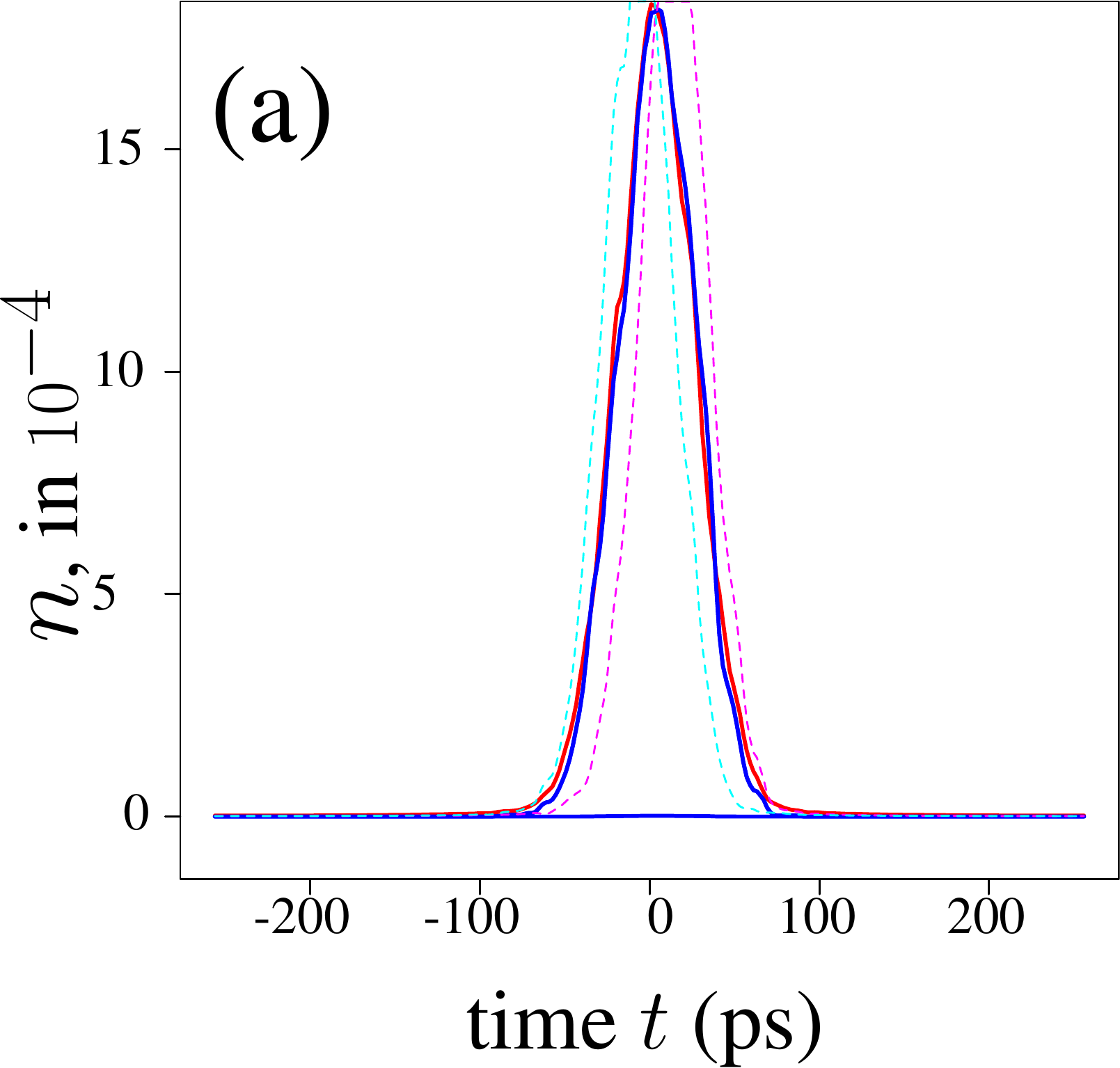}
\includegraphics[angle=0,width=0.32450\columnwidth]{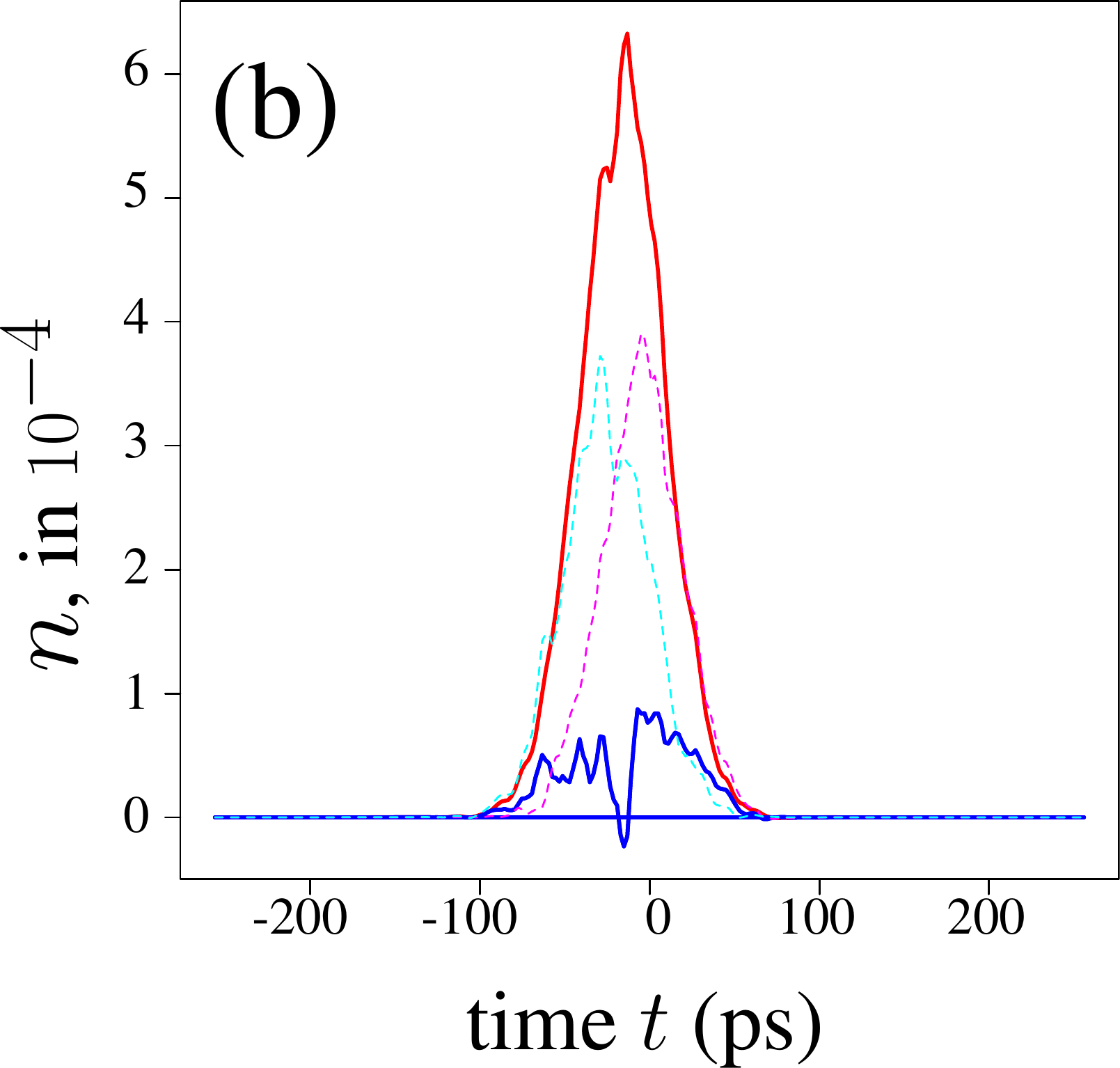}
\includegraphics[angle=0,width=0.32450\columnwidth]{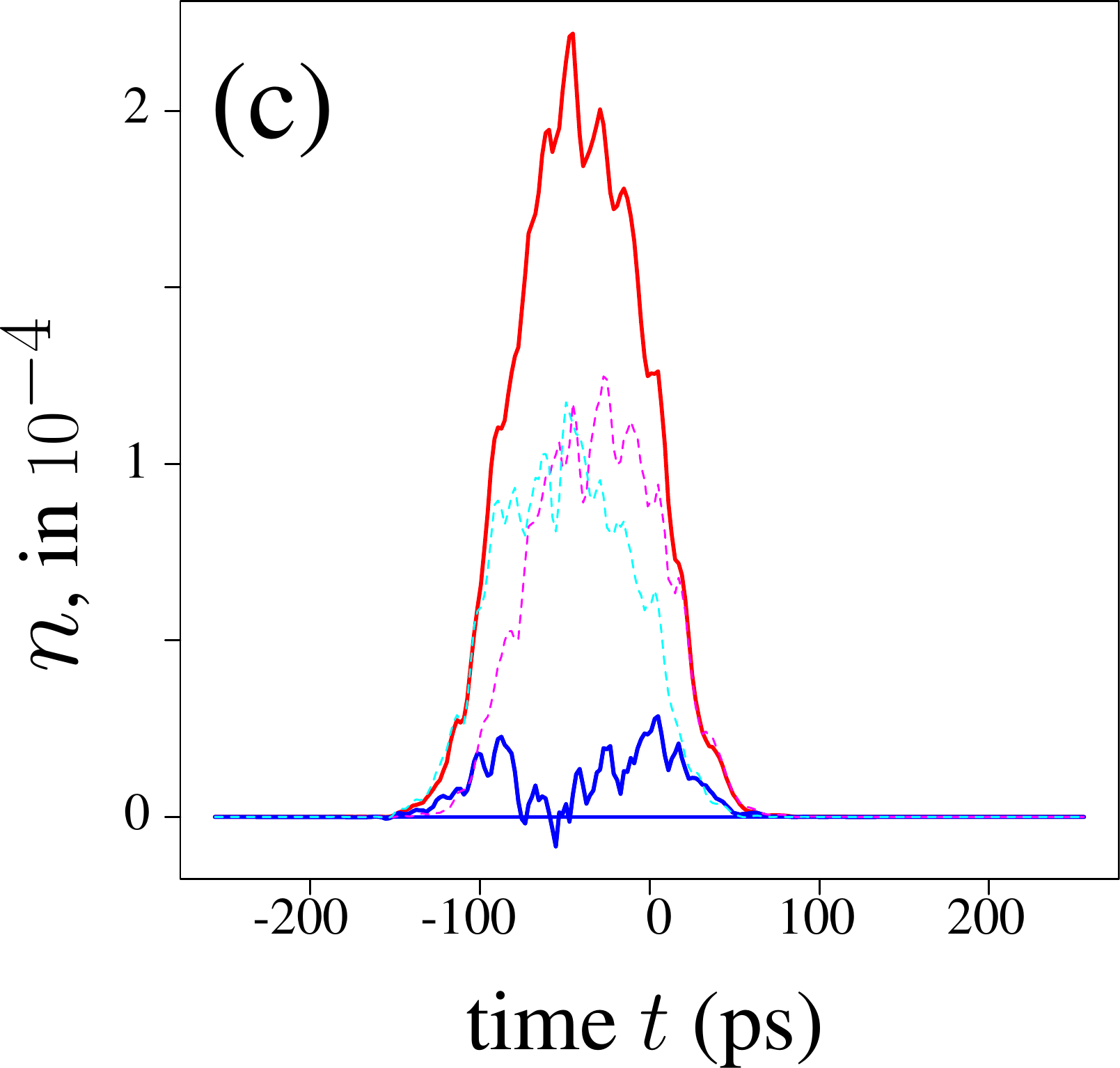}\\
\includegraphics[angle=0,width=0.32450\columnwidth]{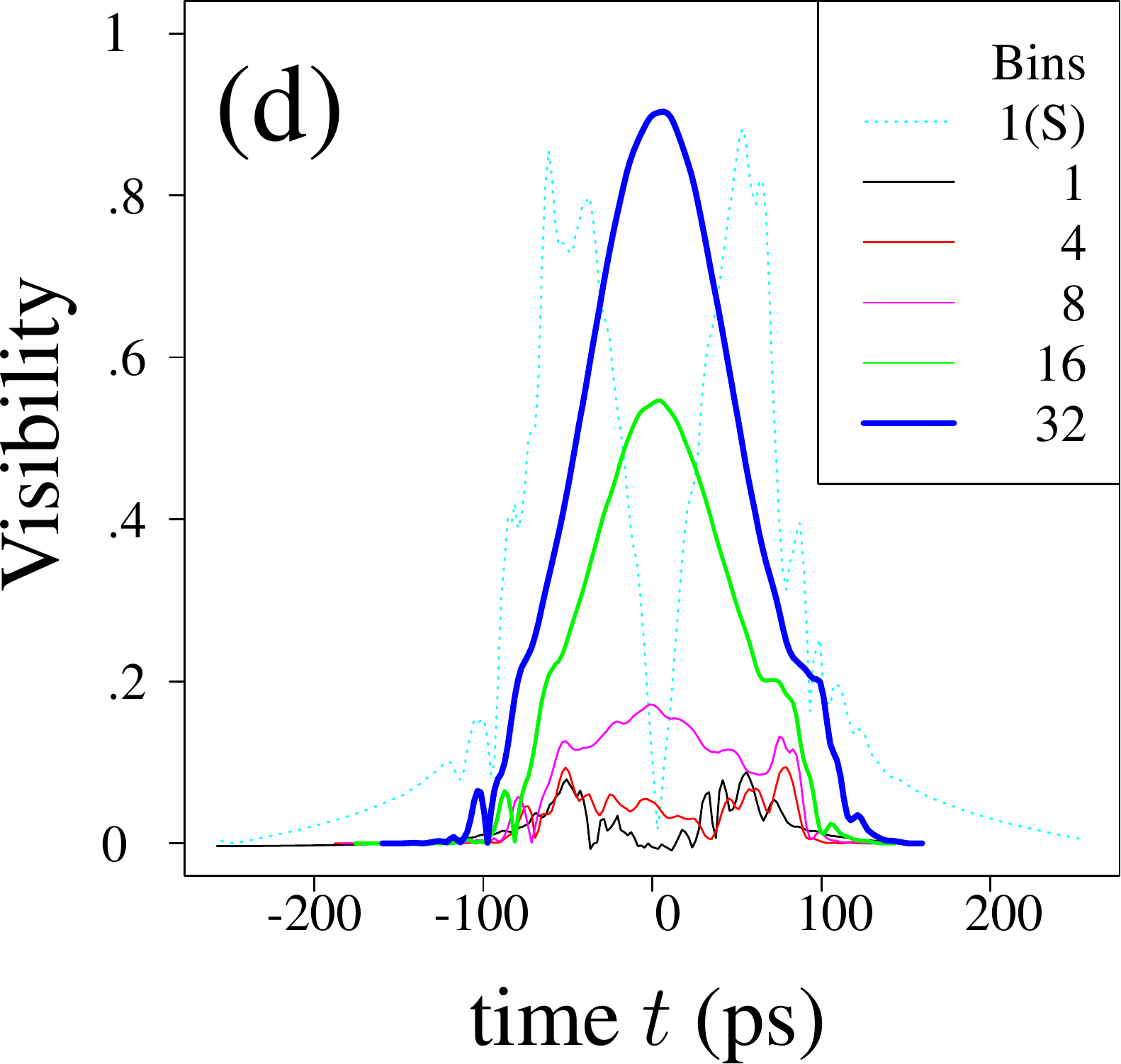}
\includegraphics[angle=0,width=0.32450\columnwidth]{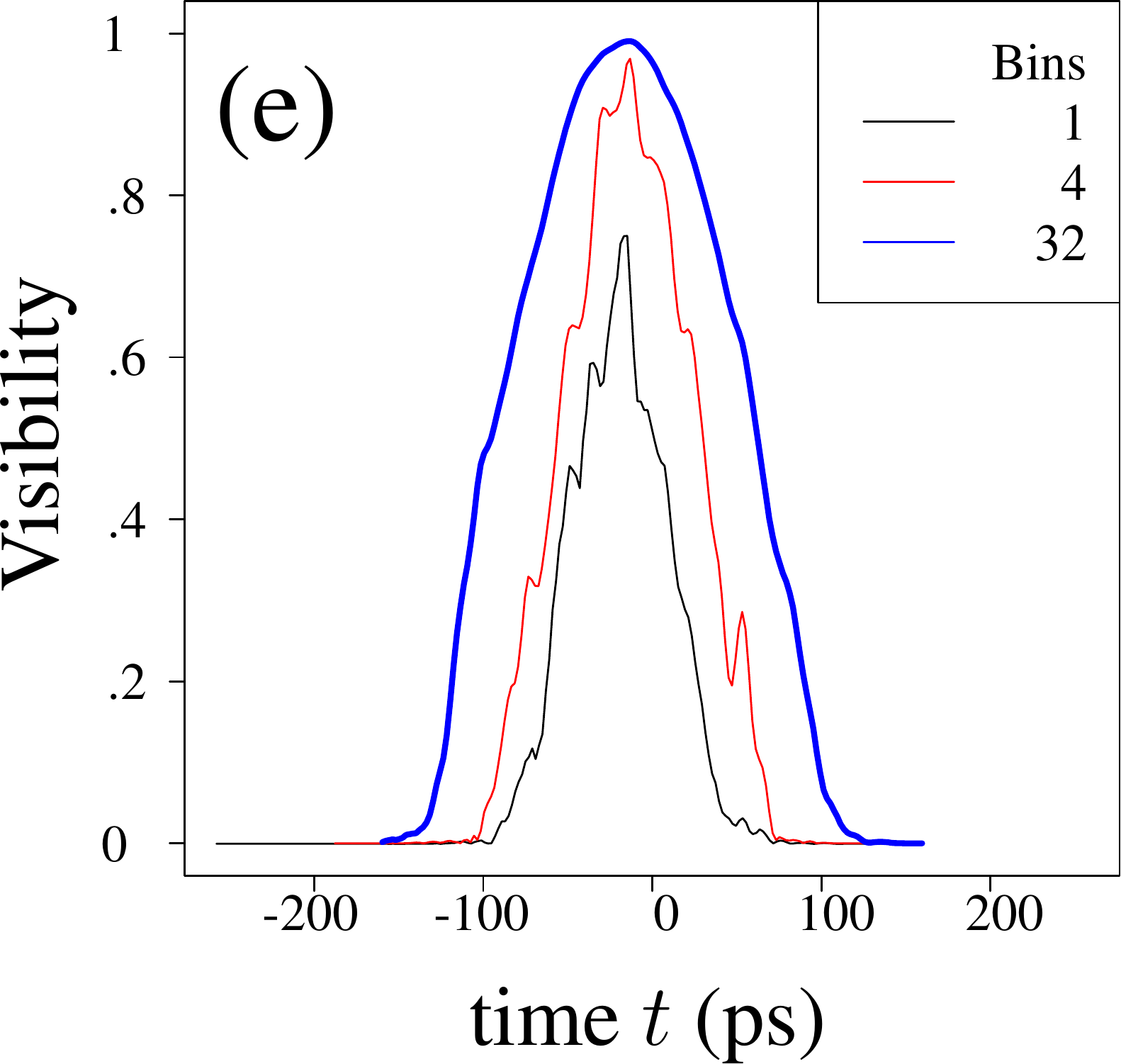}
\includegraphics[angle=0,width=0.32450\columnwidth]{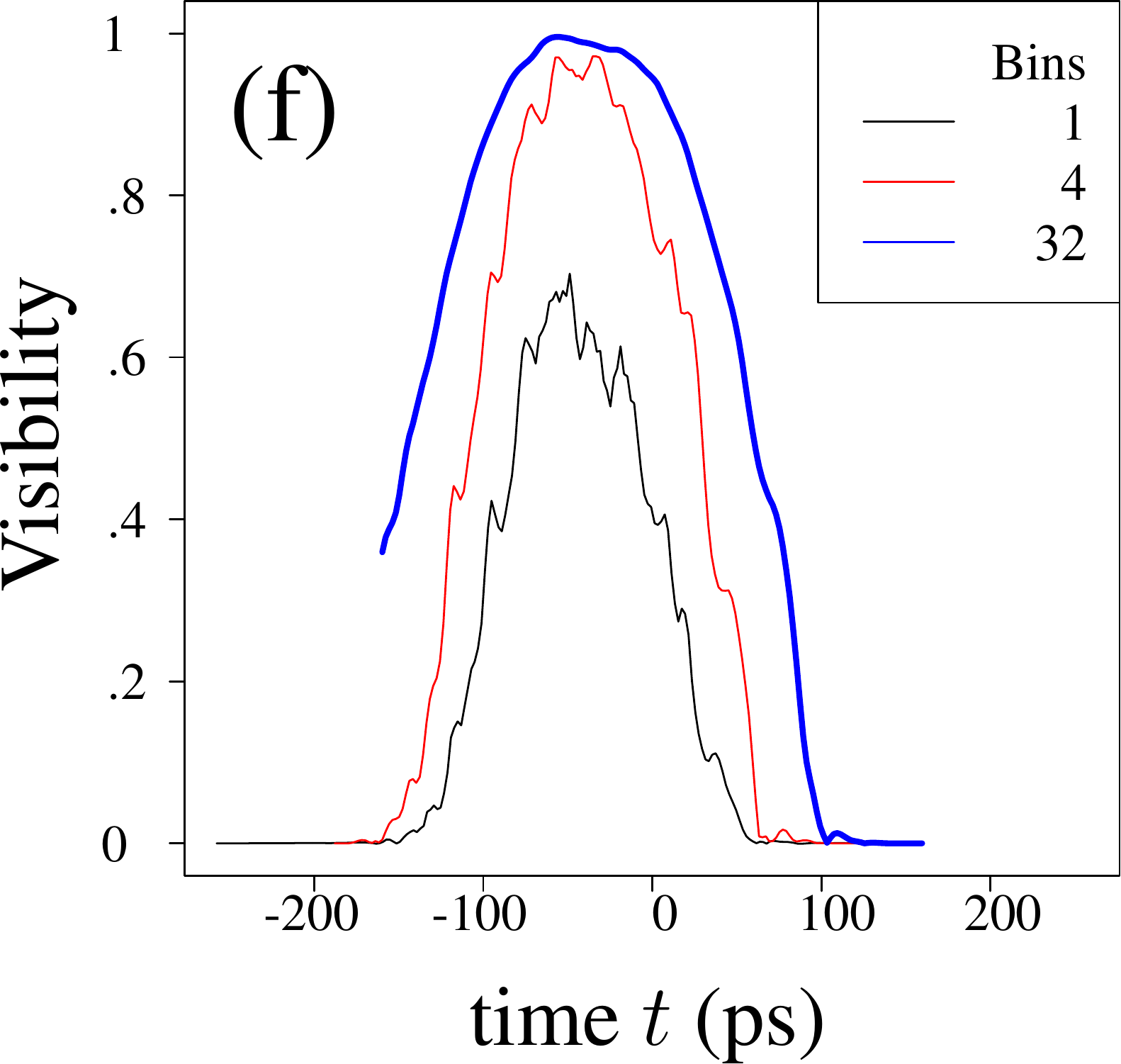}
\caption{Pulsed simulations 
 with no object ($r=0$)
 and artificially scaled ((a,d) 10\%, (b,e) 50\%)
 and true (c,f) group velocity offsets.
\textbf{~(a,b,c):}
 detected photon numbers per 2ps, $n_A, n_B$ (red and blue); 
 dotted lines are $n_{s1}, n_{s2}$ for the pre-beamsplitter
 signal fields; 
 note the change in vertical axis scale.
\textbf{~(d,e,f):}
 \cRed{detected entanglement visibilities for different time averaging}, 
  at 2ps per bin; 
 note that statistical fluctuations artificially 
 increase the computed visibility.
The 1(S) curve on (d) is a (false) visibility
 based on the pre-beamsplitter signal pulses, 
 and has a value only because of their arrival-time
 mismatch.
}
\label{fig-pulse-LastStack0607}
\end{figure}

Results for no object and standard material parameters,
 but also considering artificially reduced group velocity offsets,
 are shown on
 fig. \ref{fig-pulse-LastStack0607}\cRed{(a-f)}, 
 where they are 
 compared with full parameter results.
The displacement to negative $t$ of the visibility peak 
 is a result of the 
 group velocity walk-off of the signal field.
We see that the detected photon rates $n$ decrease 
 at larger group velocity mismatches -- 
 this is due to the increased spreading of the generated fields, 
 and hence less effective nonlinear generation.
\cRed{Despite the decrease in detected $n$, 
 we see that the reduction in detector-averaged entanglement visibility 
 is relatively minor; 
 whilst in stark contrast, 
 the drop in \emph{un-averaged} visibility is significant.}
Thus we see that
 sufficiently long averaging interval
 allows us 
 to recover most of the maximum possible visibility,
 albeit not all; 
 and as we would expect the averaging also helps reduce the 
 significant statistical variation visible on the un-averaged 
 data.  
Thus fig. \ref{fig-pulse-LastStack0607} shows that group velocity mismatches
 are not as problematic as they might at first appear,
 since the generated entanglements, 
 \cRed{however scrambled 
 they might be by the gradual nonlinear generation
 and significantly dispersive propagation},
 \emph{can} be recovered to a significant extent
 by time-averaging at the detector.

\begin{figure}[h]
\includegraphics[angle=-0,width=0.480\columnwidth]{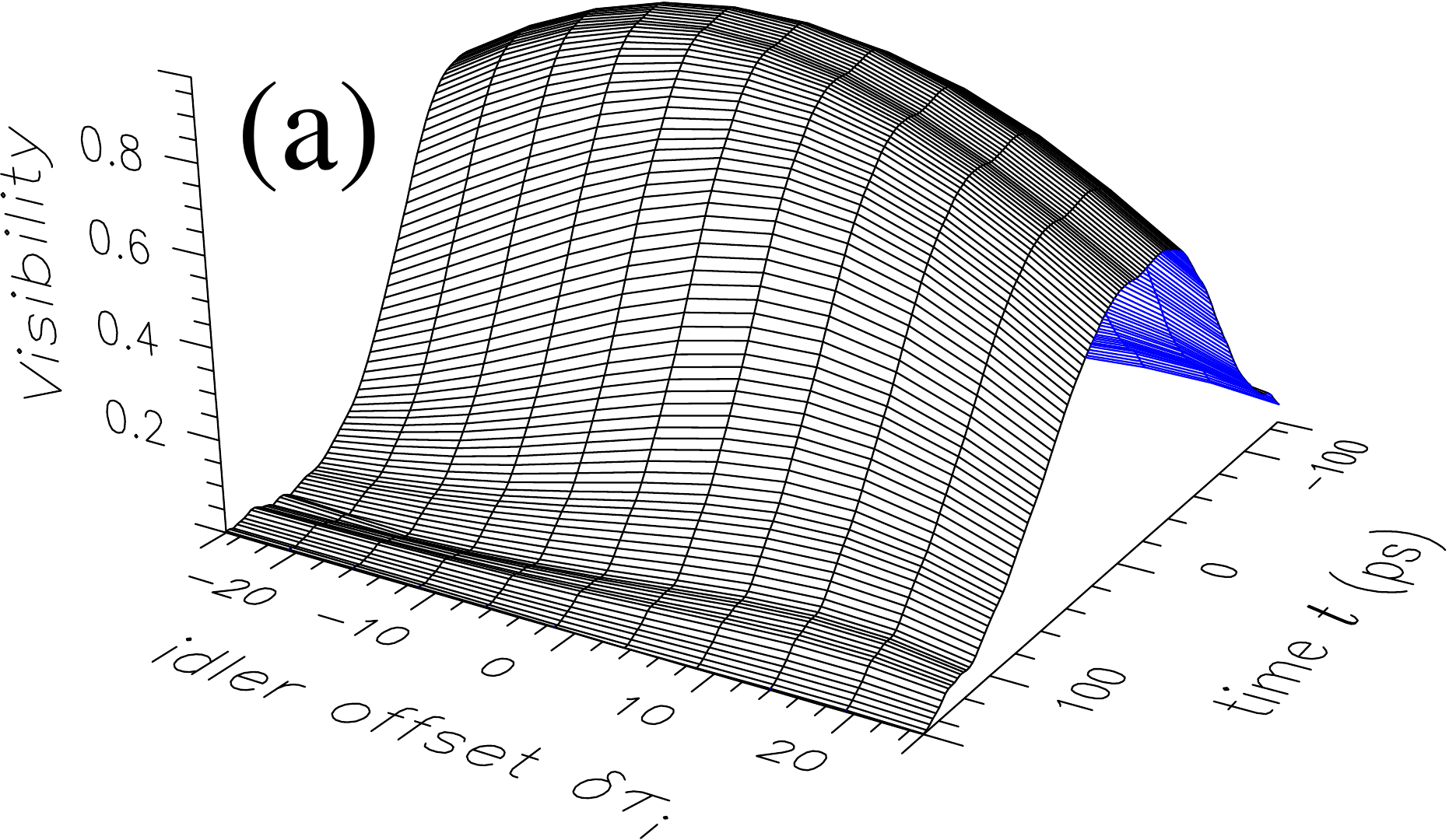}
\includegraphics[angle=-0,width=0.480\columnwidth]{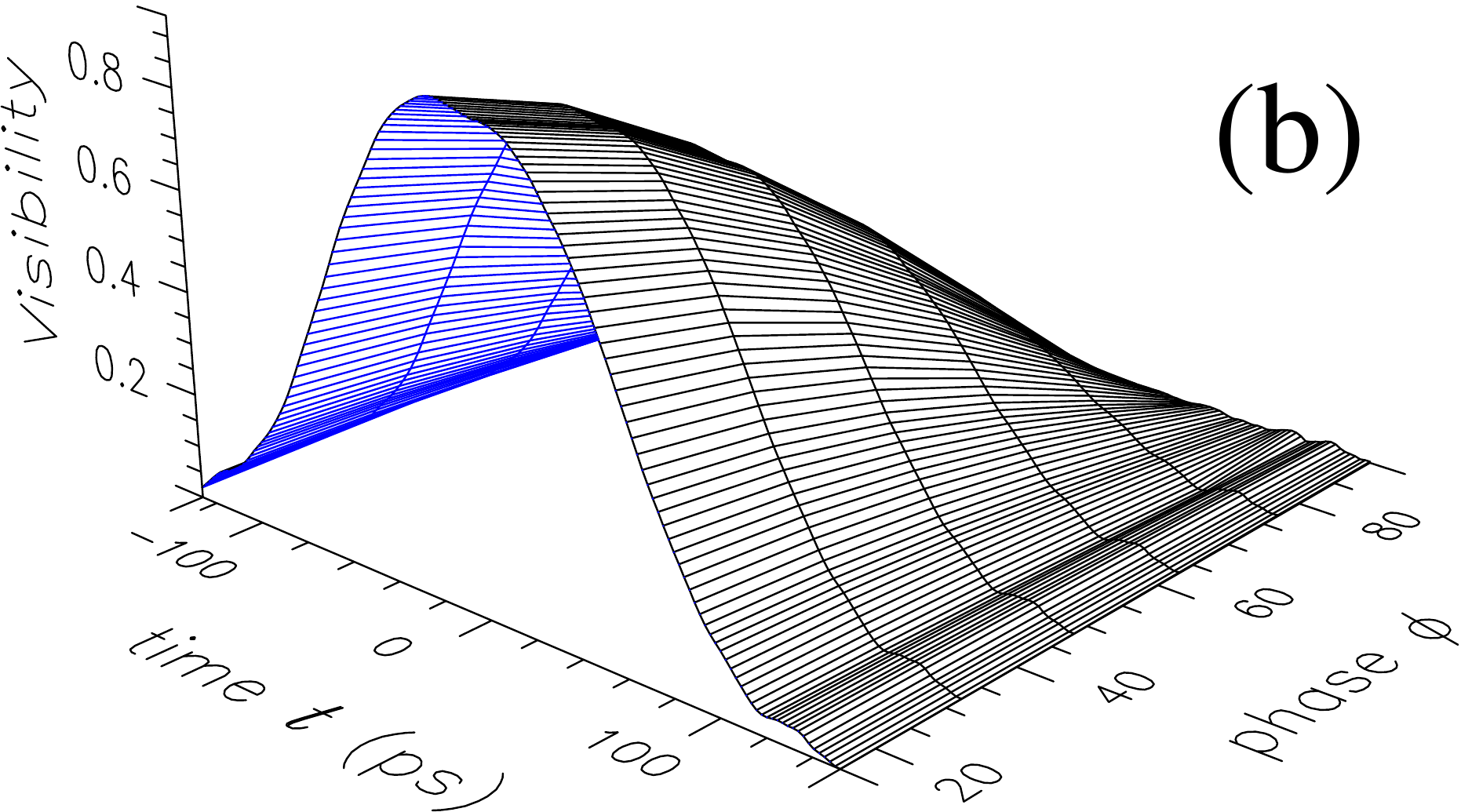}\\
\includegraphics[angle=-0,width=0.480\columnwidth]{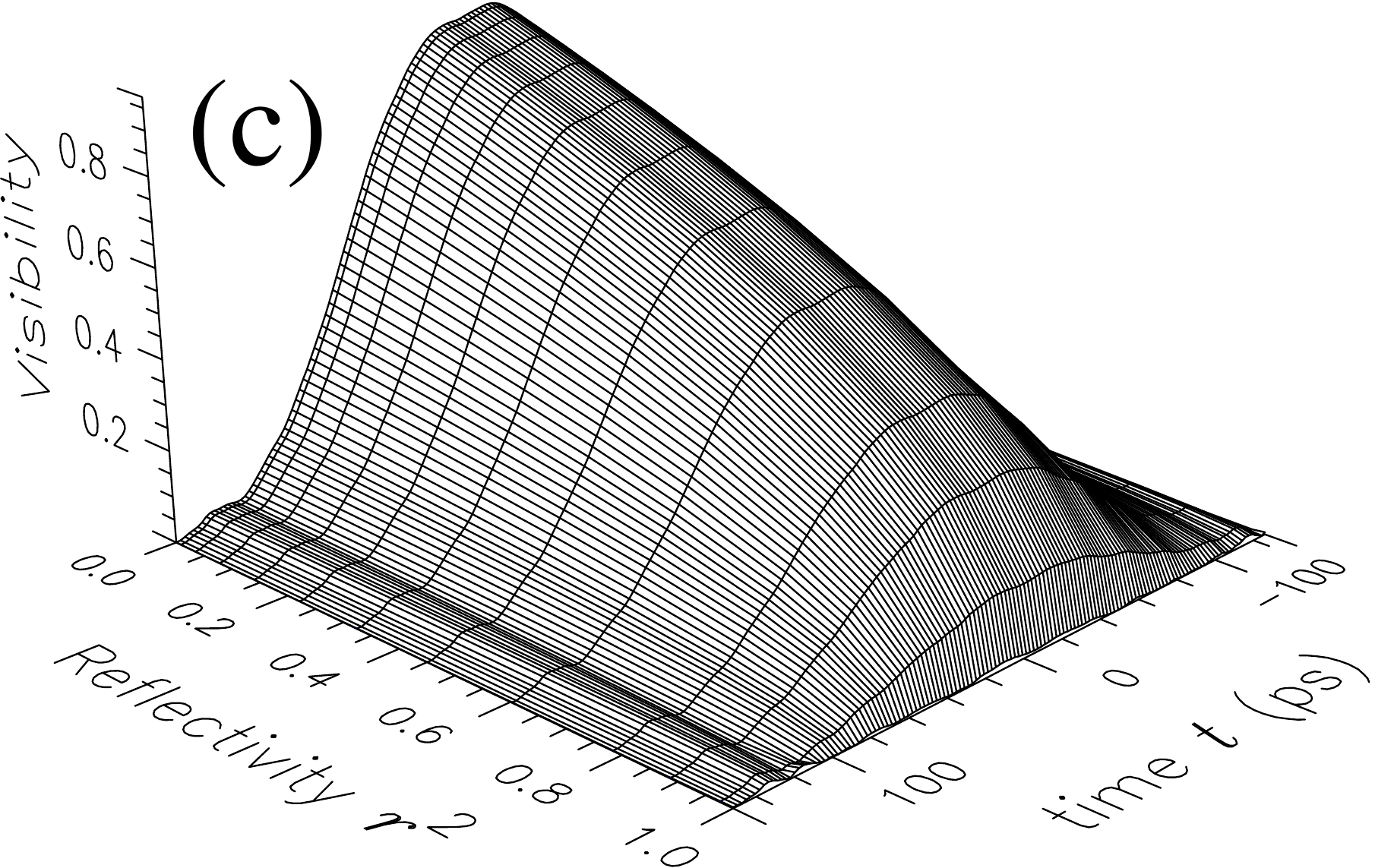}
\includegraphics[angle=-0,width=0.460\columnwidth]{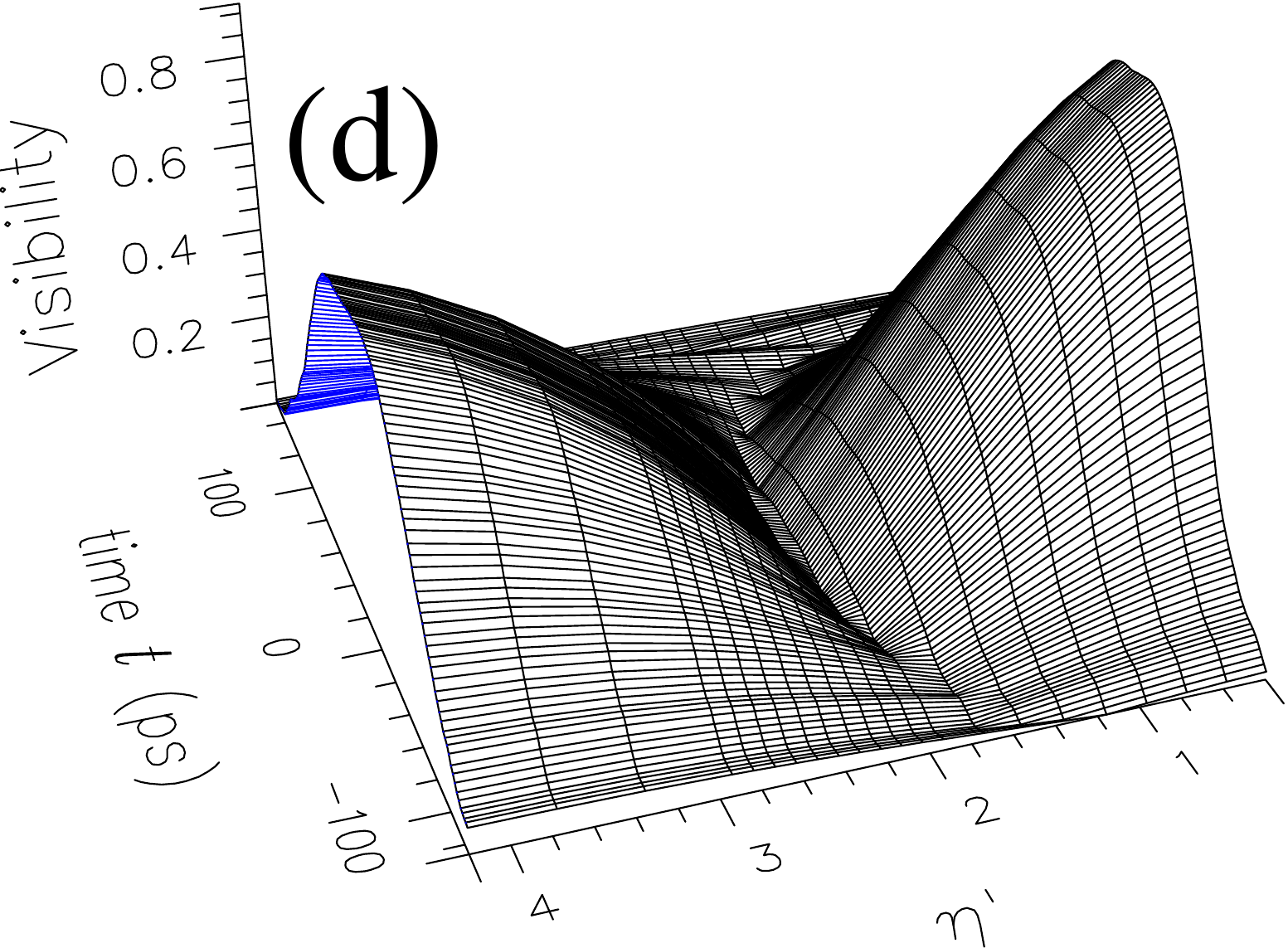}
\caption{Time-dependent entanglement visibilities for 
 pulsed simulations 
 with  32 bin (64ps) averaging
 {and a $40$ps mistiming
 at the output beamsplitter.}
In \textbf{(a)} we see 
 how the visibility profile
 changes if the idler synchonisation $\delta \tau_i$
 differs from the ideal value. 
Other frames show results for varying object properties, 
 i.e. 
 \textbf{(b)} phase,
 \textbf{(c)} reflectivity,
 and
 \textbf{(d)} interaction with
 a dynamic object
 as per \eqref{eqn-dynobject-b} -- \eqref{eqn-dynobject-backaction}.
}
\label{fig-pulse-LastStack08timed}
\end{figure}

In fig. \ref{fig-pulse-LastStack0607} the simulations 
 ensured the idler pulse arrived at the NL2 stage 
 in correct synchronization with the pump pulse.
We can see the effect of mis-timing the idler pulse
 at this point on fig. \ref{fig-pulse-LastStack08timed}(a), 
 where at least for these parameters --
 notably 40ps pulse widths --  
 the fall off in averaged visibility is gradual.
This is due to a combination of the pump pulse length ($40$ps), 
 the averaging time ($64$ps), 
 and the group velocity spreading ($\sim 80$ps).
\cRed{On fig. \ref{fig-pulse-LastStack08timed}(b,c), 
 and in broad agreement with trends in fig. \ref{fig-ideal-with-both}, 
 we see a fall-off
 of the time-dependent entanglement visibility with
 object phase depth and object reflectivity.}

%

\emph{Dynamic objects}
 \cRed{further emphasise the 
 potential role of time-dependence.} 
\cRed{Fig. \ref{fig-pulse-LastStack08timed}(d)}
 shows recovered entanglement visibility values
 for a range of interaction strengths $\eta'$.
At low $\eta'$, 
 idler field excitations are coupled into the object but not out again, 
 leading to reduced visibility.
However, 
 as the $\eta'$ increases even further,
 those excitations can also start being coupled back out, 
 leading to a partial recovery.
\cRed{This non-trivial behaviour suggests the possibility of interesting trade-offs
 when considering the imaging of dynamic objects.}

%

\emph{In conclusion,}
 we have shown how the time-averaging process
 inherent in slow detector response times 
 enables recovery of the entanglement necessary
 \cRed{quantum nonlinear interferometry,
 even when confounding effects such 
 as group velocity differences, 
 dispersion}, 
 mis-timing of pulse arrivals, 
 or objects interposed in the idler beam are allowed for.   
Indeed, 
 group velocity mismatches would normally 
 be expected to play a critical role, 
 seeing as they can rapidly de-synchronise mutually entangled
 time-slices of the light field.
This can be seen in our simulation results 
 \cRed{(e.g. in figs. \ref{fig-ideal-with-vgaveraging}(a)
  and \ref{fig-pulse-LastStack0607}(f))
 which show a significant
 improvement when changing from un-averaged to averaged visibilities}.
In contrast to visibilities, 
 which are a ratio, 
 nonlinear generation efficiencies suffer penalties
 from the effect of material dispersion regardless of averaging.
Finally,
 our linearly coupled dynamic object model
 acts as a starting point for more sophisticated interactions; 
 a feature likely to be important in systems relying on short pulses, 
 resonant behaviour, 
 and time-domain entanglement.

\emph{Acknowledgements:}
 Support 
 from the UK National Quantum Hub for
 Imaging (QUANTIC, EP/T00097X/1).


%
%



%
\section*{Supplementary Information:\\
~\\
{The surprising persistence of time-dependent quantum entanglement}}\label{S-appendix}

\setcounter{page}{1}
\pagenumbering{roman}

\noindent
\emph{Paul Kinsler, Martin W. McCall, Rupert F. Oulton, and Alex S. Clark}\\
{\small{Department of Physics, Imperial College London, United Kingdom}}\\
~



%
\subsection{Representation of entanglement}\label{S-entanglement}

In the model used here, 
 entanglement is represented as classical statistical correlations
 between complex field amplitudes
 which are capable of reproducing the off-diagonal elements
 of the density matrix --
 i.e. they can represent all the necessary quantum properties
 \cite{Drummond-G-1980jpa}.
This is possible because each field is represented 
 by two independent amplitudes $\alpha_m$ and $\alpha_m^\dagger$; 
 and although they are complex conjugate on average, 
 i.e. $<\alpha_m> = <\alpha_m^\dagger>^*$, 
 in any given trajectory making up part of the large ensemble, 
 they need not be.

By looking at the SDE's for the nonlinearity 
 \eqref{eqn-cht-daa1} -- \eqref{eqn-cht-dadp}, 
 we can see that the non-daggered and the daggered amplitudes 
 are driven by \emph{different} noises.
Thus
 both the mean photon numbers
 $<\aadi{\ksigl}\aaai{\ksigl}>$
 and
 $<\aadi{\kidlr}\aaai{\kidlr}>$
 could {even} remain nearly zero
 even whilst
 a significant quantum statistical correlation (i.e. entanglement)
 is being created between the signal and idler fields; 
 i.e. between $\aaai{\ksigl}$ and $\aadi{\ksigl}$,
 and between $\aadi{\ksigl}$ and $\aadi{\kidlr}$.

%
\subsection{Simulation statistics: reducing sampling error}\label{S-statistics}

It has long been known that that getting good simulation statistics
 with the fully quantum mechanical 
 positive-P representation can be challenging \cite{Carter-1995pra}, 
 so that 
 it is very common to resort to the much simpler, 
 but approximate, 
 truncated Wigner representation \cite{DrummondHilleryQTNO} in simulations.
In cantrast to the positive-P,
 the truncated Wigner representation is essentially 
 a semi-classical hidden variable model 
 that represents the quantum uncertainty
 as simple statistical fluctuations in the field amplitudes
 \cite{Kinsler-D-1991pra,Kinsler-D-1991pra-comment}.  
This halves the number of equations required by the simulation
 (needing only a single complex $\alpha$ rather than the 
  double $\alpha$ and $\alpha^\dagger$), 
 reducing the state space, 
 and resulting in an effective and sufficiently accurate method
 when e.g. studying quadrature squeezing \cite{Walls-1983n}, 
 and its generation using optical pulses in nonlinear materials 
 \cite{Drummond-C-1987josab,Carter-DRS-1987prl,Carter-1995pra,Corney-HDJDLA-2008pra}.

However, 
 in this work, 
 here we want to ensure an accurate representation
 of the quantum effects, 
 and so we stay with the full positive-P model
 (cf the case of quantum tunnelling \cite{Kinsler-D-1991pra,Kinsler-D-1991pra-comment}
 and nonlinearity and the quantum vacuum \cite{Kinsler-1996}).
Since the subtler effects of quantum entanglement is addressed here,
 rather than the simpler quadrature moments, 
 a truncated Wigner representation
 would not suffice, 
 since it imposes an unavoidable linkage between correlations
 and photon number.
This leaves us requiring the use of a full positive-P description
 and concommitant extremely long run times.
This is especially problematic since we may need to resolve 
 very low average photon numbers
 inside an extremely noisy background.

We address this using a hybrid strategy which 
 allows us to use
 just \emph{one} very high ensemble number $M_B$ simulation
 to get a good estimate of the background for some particular case, 
 and then adjusting this using two more simulations
 with lower ensemble numbers $M_R$ but perfectly matched 
 random noise generation.
We call the $M_B$ simulation the ``background'', 
 and the other two  the 
 ``reference'' and ``target'' simulations.
The reference simulation has identical parameters
 to the background simulations, 
 and the target simulation has the parameters
 coresponding to the particular result we are 
 interested in.
The difference between the reference and target simulations
 only depends on the differences between the system parameters, 
 with only ``second-order'' noise effects --
 resulting from how the noise influences propagation 
 differently,
 and not from different random number sequences.

When trying to evaluate a photon number $n$ 
 for some chosen target parameters,
 we proceed as follows, 
 using the usual notation where a statistical average
 is denoted $<n>$, 
 but additionally adding a subscript to denote the 
 ensemble size, 
 with $\infty$ to denote the idea infinite-ensemble case.
Each trajectory in the background or reference ensembles returns a value $n'_j$,
 and each in the target ensemble returns $n_j$.

Thus for $n'$ we can have 
 either a low sampling error, 
 or a larger sampling error, 
 as follows
~
\begin{align}
  \left< n' \right>_\infty
&\simeq
   \left< n' \right>_{M_B}
 = 
   \frac{1}{M_B} \sum_{j=1}^{M_B} n'_{j}
\label{eqn-statistics-MB}
\\
&\approx
   \left< n' \right>_{M_R}
 = 
   \frac{1}{M_R} \sum_{j=1}^{M_R} n'_{j}
,
\label{eqn-statistics-MR}
\end{align}
 with the sampling error reducing for each as 
 $M_B$ and $M_R$ are increased; 
 thus for large enough $M$ values we have
~
\begin{align}
    \left< n' \right>_{M_B}
&\approx
    \left< n' \right>_{M_R}
\\
\textrm{i.e.}\qquad
   \left< n' \right>_{M_B} -   \left< n' \right>_{M_R} 
&\approx
  0
,
\label{eqn-statistics-diffMBMR}
\end{align}
 but noting that the noise-sampling error
 in this approximate equality \eqref{eqn-statistics-diffMBMR}
 is dominated by the larger variation 
 in the smaller reference ensemble.

Similarly, 
 for $n$ we 
 we have
~
\begin{align}
  \left< n \right>_\infty
&\approx
   \left< n \right>_{M_R}
 = 
   \frac{1}{M_R} \sum_{j=1}^{M_R} n_{j}
.
\end{align}

Since \eqref{eqn-statistics-diffMBMR} should average
 to zero, 
 we can now write
~
\begin{align}
  \left< n \right>_\infty
&=
  \left< n \right>_\infty
 -
  \left< n' \right>_\infty
 +
  \left< n' \right>_\infty
\\
&\approx
   \left< n \right>_{M_R}
 -
   \left< n' \right>_{M_R}
 -
   \left< n' \right>_{M_B}
,
\end{align}
 which,
 given its dependence on both 
 $\left< n \right>_{M_R}$ and $\left< n' \right>_{M_R}$,
 would at first appear to suffer
 a larger sampling error based on the smaller $M_R$, 
 not a small sampling error based on the large $M_B$.

However, 
 the key point is that since we have \emph{exactly} matched
 the simulation noise values
 between the reference simulations of $n'_j$ 
 and target simulations of $n_j$ then their difference 
 is due to differences of the trajectory dynamics 
 between the two simulations, 
 and is only weakly dependent on the specific noise values.
Since this cancels out the bulk
 of the sampling error due to the noise
 in these smaller-ensemble simulations, 
 we can now write
~
\begin{align}
  \left< n \right>_\infty
&\simeq
   \left< n \right>_{M_R}
 -
   \left< n' \right>_{M_R}
 -
   \left< n' \right>_{M_B}
,
\end{align}
 where now it is the sampling error from
 the \emph{background} ensemble that dominates.

Thus for any parameter set, 
 we only need to do one large ensemble $M_B$ background simulation, 
 one smaller ensemble $M_R$ reference simulation
 otherwise \emph{identical} to the background one, 
 and then then many small $M_R$ target simulations which do have 
 a parameter variation compared to the background simulation.
Background parameters were of systems with perfectly transparent objects,
 and our target simulations varied only the object properties.
This meant that all the results shown on fig. \ref{fig-pulse-LastStack08timed}
 could be generated from the same background simulation, 
 a large and --
 in our case -- 
 very necessary reduction in total simulation time.

Here we typically did background simulations with $M_B$ sizes
 from $\sim 10^6$
  (fig. \ref{fig-pulse-LastStack0607}(a,d))
 to $\sim  16 \times 10^6$
  (figs. \ref{fig-pulse-LastStack0607}(c,f) and \ref{fig-pulse-LastStack08timed}), 
 and reference and target simulations
 with $M_B =  128\times 10^3$.
In the most extreme case, 
 this gave us a factor of 125 improvement in effective simulation speed, 
 as well as the crucial decrease in statistical error.
Whilst it may be possible to vary other parameters, 
 or perhaps even several parameters at once, 
 this will induce a greater divergence between 
 propagation in the reference and target systems, 
 and so affect the extent of any improvement.

\subsection{Nonlinearity}\label{S-nonlinearity}

As described in the main text, 
 the key feature of the 
 degenerate-pump spontaneous FWM
 interaction used in our simulations
 is that it produces pairs of entangled signal and idler fields (photons).
It shares this feature with the more commonly used 
 second order parametric nonlinearity \cite{DrummondHilleryQTNO} 
 (usually denoted $\chi^{(2)}$), 
 which generates the same kind of entangled pairs
 but from single pump photons, not pairs.

Although this difference is not trivial, 
 the SDE equations for a $\chi^{(2)}$ nonlinearity are similar in form,
 although with no quantum noise term applied to the pump field.
To test what differences might appear between the two models, 
 some simulations were also done with this $\chi^{(2)}$ model
 and system parameters rescaled to match the nonlinear effects; 
 the results were remarkably similar in character, 
 indicating that our conclusions are not specific to the FWM model 
 we present here.

Note that 
 the nonlinear response is treated here as if it were instantaneous.
This is a reasonable approximation 
 since typical nonlinear response times in dielectrics 
 are of the order of femtoseconds or less
 \cite{Das-BGBHSRHVNM-2013epjconf}.

%
\subsection{Evolution: from density matrix to SDEs}\label{S-fwm-evolution}

\newcommand\qpartial[1]{\frac{\partial}{\partial{#1}}}

Since to our knowledge the positive-P Fokker-Planck equation
 for the degenerate-pump FWM interaction Hamiltonian 
 are not currently in the literature, 
 so we now sumarize the derivation.
This interaction Hamiltonian \eqref{eqn-dfwm-hamiltonian}
 gives a contribution to the density matrix evolution
~
\begin{align}
  \dot{\rho}_{\textup{int}}
&=
  \frac{\imath}
       {\hbarr}
  \left[
   \hat{H}_{\textup{int}}, \rho
  \right]
.
\end{align}

The expansion of the density matrix in terms of coherent states, 
 albeit for just a single mode,
 is
~
\begin{align}
  \rho
&=
  \int
    \frac{  \left| \alpha \right> \left< \alpha^\dagger \right| }
         {  \left< \alpha | \alpha^\dagger \right> }
    P(\alpha, \alpha^\dagger; t)
  ~
  d^2\alpha ~ d^2\alpha^\dagger
.
\end{align}
Using the standard positive-P operator correspondences, 
 we can convert the density matrix dynamics
 into a dynamics for the corresponding
 positive-P distribution function
 \cite{Drummond-G-1980jpa,GardinerHSM,DrummondHilleryQTNO}.
 which is a function of many complex coherent state amplitudes.
The resulting Fokker-Planck equation for this $P(...; t)$
 is
 ~
\begin{align}
  \partial_t
    P(...; t)
&=
  \mathscr{L}_{\textup{int}}
    P(...; t)
.
\end{align}

The derivative operators
 defining this part of the dynamics for the positive-P distribution $P$ 
 is as follows.
The first two lines result from the 
 first commutator term $\hat{H}_{\textup{int}} \rho$, 
 and the last two lines from $-\rho \hat{H}_{\textup{int}}$, 
 are
~
\begin{align}
  \mathscr{L}_{\textup{int}}
&=
 -
 {\knlc}
  \left(
    \aadi{\kpump} + \qpartial{\aaai{\kpump}}
  \right)^2
  \aaai{\ksigl}
  \aaai{\kidlr}
\nonumber
\\
&\quad
 + 
  {\knlc}^*
  \aaai{\kpump} \aaai{\kpump}
  \left(
    \aadi{\ksigl} + \qpartial{\aaai{\ksigl}}
  \right)
  \left(
    \aadi{\kidlr} + \qpartial{\aaai{\kidlr}}
  \right)
\nonumber
\\
&\quad\quad
 +
  {\knlc}
  \aadi{\kpump} \aadi{\kpump}
  \left(
    \aaai{\ksigl} + \qpartial{\aadi{\ksigl}}
  \right)
  \left(
    \aaai{\kidlr} + \qpartial{\aadi{\kidlr}}
  \right)
\nonumber
\\
&\quad\quad\quad
 -
  {\knlc}^*
  \left(
    \aaai{\kpump} + \qpartial{\aadi{\kpump}}
  \right)^2
  \aadi{\ksigl}
  \aadi{\kidlr}
.
\end{align}


Expanded we have, 
~
\begin{align}
~
&
 -
 {\knlc}
  \left\{
    \aadi{\kpump} \aadi{\kpump} 
   +
    \aadi{\kpump} \qpartial{\aaai{\kpump}}
   +
    \qpartial{\aaai{\kpump}} \aadi{\kpump} 
   +
    \qpartial{\aaai{\kpump}} \qpartial{\aaai{\kpump}}
  \right\}
  \aaai{\ksigl}
  \aaai{\kidlr}
\nonumber
\\
&\quad
 + 
 {\knlc}^*
  \aaai{\kpump} \aaai{\kpump}
  \left\{
    \aadi{\ksigl} \aadi{\kidlr}
   +
    \aadi{\ksigl} \qpartial{\aaai{\kidlr}}
   +
    \qpartial{\aaai{\ksigl}} \aadi{\kidlr}
   +
    \qpartial{\aaai{\ksigl}} \qpartial{\aaai{\kidlr}}
  \right\}
\nonumber
\\
&\quad\quad
 +
 {\knlc}
  \aadi{\kpump} \aadi{\kpump}
  \left\{
    \aaai{\ksigl} \aaai{\kidlr}
   +
    \aaai{\ksigl} \qpartial{\aadi{\kidlr}}
   +
    \qpartial{\aadi{\ksigl}} \aaai{\kidlr}
   +
    \qpartial{\aadi{\ksigl}} \qpartial{\aadi{\kidlr}}
  \right\}
\nonumber
\\
&\quad\quad\quad
 -
 {\knlc}^*
  \left\{
    \aaai{\kpump} \aaai{\kpump}
   +
    \qpartial{\aadi{\kpump}} \aaai{\kpump} 
   +
    \aaai{\kpump} \qpartial{\aadi{\kpump}}
   +
    \qpartial{\aadi{\kpump}} \qpartial{\aadi{\kpump}}
  \right\}
  \aadi{\ksigl}
  \aadi{\kidlr}
.
\end{align}

Now all the terms without derivatives cancel, 
 so that 
~
\begin{align}
~
&
 -
 {\knlc}
  \left\{
    \aadi{\kpump} \qpartial{\aaai{\kpump}}
   +
    \qpartial{\aaai{\kpump}} \aadi{\kpump} 
   +
    \qpartial{\aaai{\kpump}} \qpartial{\aaai{\kpump}}
  \right\}
  \aaai{\ksigl}
  \aaai{\kidlr}
\nonumber
\\
&\quad
 + 
 {\knlc}^*
  \aaai{\kpump} \aaai{\kpump}
  \left\{
    \aadi{\ksigl} \qpartial{\aaai{\kidlr}}
   +
    \qpartial{\aaai{\ksigl}} \aadi{\kidlr}
   +
    \qpartial{\aaai{\ksigl}} \qpartial{\aaai{\kidlr}}
  \right\}
\nonumber
\\
&\quad\quad
 + 
 {\knlc}
  \aadi{\kpump} \aadi{\kpump}
  \left\{
    \aaai{\ksigl} \qpartial{\aadi{\kidlr}}
   +
    \qpartial{\aadi{\ksigl}} \aaai{\kidlr}
   +
    \qpartial{\aadi{\ksigl}} \qpartial{\aadi{\kidlr}}
  \right\}
\nonumber
\\
&\quad\quad\quad
 -
 {\knlc}^*
  \left\{
    \qpartial{\aadi{\kpump}} \aaai{\kpump} 
   +
    \aaai{\kpump} \qpartial{\aadi{\kpump}}
   +
    \qpartial{\aadi{\kpump}} \qpartial{\aadi{\kpump}}
  \right\}
  \aadi{\ksigl}
  \aadi{\kidlr}
.
\end{align}


Now we organise the terms.
Collecting the first derivative terms, 
 which are deterministic ``drift'' terms, 
 results in 
~
\begin{align}
~
&
 -
  \qpartial{\aaai{\kpump}}
  \left[
    2 {\knlc} \aadi{\kpump} \aaai{\ksigl} \aaai{\kidlr}
  \right]
 -
  \qpartial{\aadi{\kpump}}
  \left[
    2 {\knlc}^* \aaai{\kpump} \aadi{\ksigl} \aadi{\kidlr}
  \right]
\nonumber
\\
&\quad
 + 
  \qpartial{\aaai{\ksigl}}
  \left[
    {\knlc}^* \aaai{\kpump} \aaai{\kpump} \aadi{\kidlr} 
  \right]
 +
  \qpartial{\aadi{\ksigl}}
  \left[
    {\knlc} \aadi{\kpump} \aadi{\kpump} \aaai{\kidlr} 
  \right]
\nonumber
\\
&\quad\quad
 +
  \qpartial{\aaai{\kidlr}}
  \left[
    {\knlc}^* \aaai{\kpump} \aaai{\kpump} \aadi{\ksigl} 
  \right]
 +
  \qpartial{\aadi{\kidlr}}
  \left[
    {\knlc} \aadi{\kpump} \aadi{\kpump} \aaai{\ksigl} 
  \right]
.
\end{align}
These can be immediately converted into SDE drift terms
 where the leading derivative supplies the ``which field'' information
 and the argument (in square brackets ``[...]'')
 supplies the change in that field.

Collecting the second derivative terms, 
 which are noise-like diffusion terms, 
 results in 
~
\begin{align}
~
&
 -
  \frac{1}{2}
  \qpartial{\aaai{\kpump}}
  \qpartial{\aaai{\kpump}}
  \left\{
    2 {\knlc} \aaai{\ksigl} \aaai{\kidlr}
  \right\}
 -
  \frac{1}{2}
  \qpartial{\aadi{\kpump}}
  \qpartial{\aadi{\kpump}}
  \left\{
    2 {\knlc}^* \aadi{\ksigl} \aadi{\kidlr}
  \right\}
\nonumber
\\
&\quad
 + 
  \frac{1}{2}
  \qpartial{\aaai{\ksigl}} 
  \qpartial{\aaai{\kidlr}}
  \left\{
    2 {\knlc}^* \aaai{\kpump} \aaai{\kpump}
  \right\}
 +
  \frac{1}{2}
  \qpartial{\aadi{\ksigl}} 
  \qpartial{\aadi{\kidlr}}
  \left\{
    2 {\knlc} \aadi{\kpump} \aadi{\kpump}
  \right\}
.
\end{align}
These inform us as to the noise terms 
 and their correlations that will appear 
 in a SDE equivalent picture; 
 the noise amplitudes being the square root of the argument in braces.

These Fokker-Planck equation terms 
 can be converted into temporal SDEs 
 using standard techniques
 \cite{Drummond-G-1980jpa,GardinerHSM,DrummondHilleryQTNO},
 and by then transforming them into a co-moving frame
 \cite{Carter-1995pra}, 
 we can arrive at the set of spatially propagated SDE's 
 \eqref{eqn-cht-daa1} --
 \eqref{eqn-cht-dadp}
 used in the simulation model.
Although reasonable in the case of weak dispersion, 
 as is the case here, 
 in general
 the conversion of a material's dispersive response
 between the temporally propagated and spatially propagated domains
 is not straightforward
 \cite{Kinsler-2018jo-d2owe,Kinsler-2018jpc}.

%
\subsection{Power dependence}\label{S-power}

As already stated, 
 we have to run the simulations at a much higher pump power
 than in our nominal experimental target \cite{Pearce-POC-2020apl}
 so as to get good simulation statistics 
 (i.e. at least $\sim 10^6$ higher).
This means that we 
 rely on the scaling behaviour of the FWM terms
 in \eqref{eqn-cht-daa1} to \eqref{eqn-cht-dad2},
 the weakness of the nonlinearity, 
 and the simulation's lack of any SPM term to nevertheless
 still give representative results.

However, 
 it is important to remember that 
 we are not here attempting some exact simulation 
 of a QNI experiment based on Pearce et al \cite{Pearce-POC-2020apl}, 
 but instead we are using it as a representative scheme 
 to test the generation and recovery of entanglement information
 under the influence of material dispersion.
As a result, 
 the specific pump power used is not of direct relevance
 to our conclusions
 regarding the recovery of useful entanglement information
 as a result of the time-averaging at the detector.

\end{document}